\pdfoutput=1
\documentclass[12pt]{article}
\usepackage[utf8]{inputenc}
\usepackage[letterpaper]{geometry}
\usepackage{booktabs}
\usepackage{textcomp}
\usepackage{mathrsfs}
\usepackage{amsmath}
\usepackage{amssymb}
\usepackage{undertilde}
\usepackage{graphicx}
\usepackage{setspace}
\usepackage[authoryear]{natbib}
\makeatletter

\providecommand{\tabularnewline}{\\}

\usepackage{enumerate}
\usepackage{url}

\usepackage{geometry}
\usepackage{setspace}
\usepackage{epsfig}
\usepackage{amsfonts}
\usepackage{amsthm}
\usepackage{bbm}

\newcommand{\indicator}{\mathbbm{1}}

\title{Bayesian analysis of multifidelity computer models with local features
and non-nested experimental designs: Application to the WRF model}
\author{Bledar A. Konomi \thanks{The two authors contributed equally to this work. 
Corresponding authors: Bledar A. Konomi (alex.konomi@uc.edu) and 
Georgios Karagiannis (georgios.karagiannis@durham.ac.uk).}
\\
 Department of Mathematical Sciences, University of Cincinnati, USA
\\
 and \\
 Georgios Karagiannis \footnotemark[1] \\
 Department of Mathematical Sciences, Durham University, UK}

\@ifundefined{showcaptionsetup}{}{%
 \PassOptionsToPackage{caption=false}{subfig}}
\usepackage{subfig}
\makeatother

\begin{document}
\begin{singlespace}
\maketitle 
\end{singlespace}

\begin{singlespace}
\begin{abstract}
\noindent We propose a multi-fidelity Bayesian emulator for the analysis
of the Weather Research and Forecasting (WRF) model when the available
simulations are not generated based on hierarchically nested experimental
design. The proposed procedure, called Augmented Bayesian Treed Co-Kriging,
extends the scope of co-kriging in two major ways. We introduce a
binary treed partition latent process in the multifidelity setting
to account for non-stationary and potential discontinuities in the
model outputs at different fidelity levels. Moreover, we introduce
an efficient imputation mechanism which allows the practical implementation
of co-kriging when the experimental design is non-hierarchically nested
by enabling the specification of semi-conjugate priors. Our imputation
strategy allows the design of an efficient RJ-MCMC implementation
that involves collapsed blocks and direct simulation from conditional
distributions. We develop the Monte Carlo recursive emulator which
provides a Monte Carlo proxy for the full predictive distribution
of the model output at each fidelity level, in a computationally feasible
manner. The performance of our method is demonstrated on a benchmark
example, and compared against existing methods. The proposed method
is used for the analysis of a large-scale climate modeling application
which involves the WRF model.

\end{abstract}
\end{singlespace}

\section{Introduction}

Understanding the behavior as well as the underlying mechanisms of
real systems such as physical procedures is central to many applications
such as weather forecasting. Direct investigation of the real system
is often impossible due to limited resources, and hence it is simulated
by computer models aiming at reproducing the real system's behavior
with high accuracy. Our case study involves an expensive computer
model which requires a significant amount of resources to perform
a single run; and hence, only a limited number of simulations can
be performed. Gaussian process (GP) regression models \citep{Sacks1989}
are statistical models that allow the emulation of the computer model
output by using only a few runs of the computer model.

Computer models are often able to run at different levels of fidelity,
sophistication, or resolution. As high fidelity runs are usually more
expensive, collecting data by simulating the model at different fidelity
levels is preferred for a given budget of resources. Statistical inference
is preferable to be made against the whole simulated data-set, and
thus account for across fidelity level dependence, rather than against
simulation data-sets associated with individual fidelity levels \citep{KennedyOhagan2000}.
Assume there are available $S$ deterministic computer models $\{\mathscr{C}_{t}\}_{t=1}^{S}$
aiming at simulating the same real system. The models are ordered
by ascending fidelity level $t$. Let $y_{t}(x):\mathcal{X}\rightarrow\mathbb{R}$
denote the output function of the computer model $\mathscr{C}_{t}$
with respect to a $m$-dimensional input $x\in\mathcal{X}$. Autoregressive
co-kriging assumes 
\begin{align}
y_{t}(x) & =\xi_{t-1}(x)y_{t-1}(x)+\delta_{t}(x)\qquad\text{for}\:x\in\mathcal{X},\,t=2,...,S\label{eq:kenedyohagandynamiclink-1}
\end{align}
where $y_{t-1}(x)$ , $\delta_{t}(x)$, $\xi_{t-1}(x)$ are independent
unknown functions a priori modeled as Gaussian processes. Here, $\delta_{t}(\cdot)$
is the location discrepancy function (representing a local adjustment
from $\mathscr{C}_{t-1}$ to $\mathscr{C}_{t}$), and $\xi_{t}(\cdot)$
is the scale discrepancy (representing a scale change from $\mathscr{C}_{t-1}$
to $\mathscr{C}_{t}$ for $t=1,...,S$). Discrepancy terms, $\{\delta_{t}(\cdot)\}$
and $\{\xi_{t}(\cdot)\}$, can be thought of as accounting for `missing'
or `misrepresented' physical properties in the lower fidelity computer
model $\mathscr{C}_{t-1}$ with respect to the higher one $\mathscr{C}_{t}$.
Model \eqref{eq:kenedyohagandynamiclink-1} is induced by the Markovian
condition $\text{cov}(y_{t}(x),y_{t-1}(x')|y_{t-1}(x))=0$; i.e, there
is nothing more to learn about $y_{t}(x)$ from $y_{t-1}(x')$ for
any $x'\neq x$ given $y_{t-1}(x)$ is known.

A number of important variations of the autoregressive co-kriging
have been proposed. \citet{QianWu2008} considered the scale discrepancy
as a function of the input space by casting it as a GP. In practice,
this approach is applicable to problems with only two fidelity levels,
as the computational overhead caused by using more fidelity levels
is increased dramatically. \citet{Gratiet_SIAM_UQ2013,Gratiet_Garnier2014}
modeled the scale discrepancy as an expansion of bases defined on
the inputs, and presented conditional conjugate priors which lead
to standard conditional posterior distributions for the unknown coefficients
of the expansion. However, casting the scale discrepancy as a basis
expansion may require an undesirably large number of bases in order
to describe small scale discrepancies; while it cannot represent discontinuities
and sudden changes. Furthermore, this may aggravate non-identifiability
between the scale and additive discrepancies. \citet{perdikaris2015}
proposed a machine learning framework, which uses sparse precision
matrices of Gaussian-Markov random fields introduced by \citet{Lindgrenrue2011}.
This facilitates computations that leverage on the sparsity of the
resulting discrete operators. \citet{perdikaris2017nonlinear} relaxed
the auto-regressive structure by using deep learning ideas, however
the computational demands to train the model are significantly increased.
The aforementioned developments require hierarchically nested experimental
designs for computational reasons, otherwise the computational demands
become impractical. This constraint prevents their practical implementation
on a number of important real problems where the available data-set
is not based on such nested designs.

Our case study and motivation is a real world application that involves
the Weather Research and Forecasting (WRF) regional climate model
\citep{SkamarockKlempDudhiaGillBarkerDudaHuangWangPowers2008}. WRF
is an expensive computer model that allows the use of different resolutions
leading to different fidelity levels. We consider the WRF with the
Rapid Radiative Transfer Model for General Circulation Model \citep{PincusBarkerMorcrette2003},
with the Kain-Fritsch convective parametrisation scheme (KF CPS) \citep{Kain2004},
and with five input parameters, while we are interested in the average
precipitation as an output. The available simulations were generated
by running WRF at two resolution levels, $12.5$km and $25$km grid
spacing. The fidelity of the simulations increases when the grid spacing
gets finer. The available simulations have not been generated based
on a hierarchically nested design, while it is not possible to re-run
the expensive computer model in our facilities and generate simulations
based on such a design due to the high computational cost required.
The aforesaid co-kriging methods cannot be implemented directly due
to the lack of nested design, and hence new developments are required.
We are interested in designing an accurate emulator that aggregates
all the available simulations as well as represents features of the
WRF. Previous research in \citep{YanQianLinLeungYangFu2014,YangQianLinLeungZhang2012}
suggested that discrepancies between the two levels may depend on
the five inputs of the KF CPS. Interest also lies in better understanding
how different grid spacing affects the discrepancies in WRF with respect
to the input parameters. Existing co-kriging methods do not model/account
for such behaviors, thus suitable extensions must be introduced.

We propose the Augmented Bayesian Treed co-kriging (ABTCK); a fully
Bayesian method for building multifidelity emulators of computer models
that extends the scope of co-kriging mainly in two ways. The proposed
method is able to address applications where the available training
data-set has not necessarily been generated according to a hierarchically
nested experimental design. To achieve this, we introduce a suitable
imputation mechanism that augments the original data-set with uncertain
quantities which can be thought of as missing data from a hypothetical
complete data-set generated based on an hierarchically nested design.
The proposed imputation allows the specification of conditional conjugate
priors, and analytic integration of a large number of dimensions from
the posterior. Moreover, our method is able to account for non-stationary,
and possible discontinuities. This is achieved by suitably specifying
the statistical model as a combination of computationally convenient
and simple GP models by using a binary treed partition which a priori
follows a process similar to \citep{Chipman98,Gramacy08}. The additional
flexibility of the proposed model aims at producing more accurate
predictions as well as providing an insight of the model discrepancies.
To facilitate inference, we propose a reversible jump Markov chain
Monte Carlo (RJ-MCMC) implementation, tailored to the proposed model,
that involves an efficient MCMC sampler which operates on the joint
space of the missing data and the parameters, and consists of collapsed
blocks. Due to the augmentation, the MCMC loop consists of local RJ
updates operating on a lower dimensional state space and producing
more acceptable proposals, and a block simulating the missing data
directly from the conditionals. Finally, we propose the Monte Carlo
recursive emulator, as an alternative to those in \citep{KennedyOhagan2000,Gratiet_Garnier2014,Gratiet_SIAM_UQ2013},
which is able to provide fully Bayesian posterior predictive inference
even with non-nested designs while keeping the computational cost
lower than the others.

The rest of the paper is organized as follows. In Section \ref{sec:The-Augmented-Bayesian},
we present the proposed procedure; in Section \ref{sec:simulations},
we provide numerical comparisons with other methods; and in Section
\ref{sec:Application}, we implement our procedure for the analysis
of the WRF model. Conclusions are presented in Section \ref{sec:Conclusions-and-further}.

\section{The Augmented Bayesian Treed co-Kriging\label{sec:The-Augmented-Bayesian}}


We describe the development of our Augmented Bayesian treed co-kriging
model (ABTCK) which extends the scope of co-kriging to applications
with non-nested designs and/or non-stationary model outputs.

\subsection{Treed auto-regressive co-kriging \label{subsec:Extending-the-model}}

To account for non-stationarity we consider an known partition $\{\mathcal{X}_{k}\}_{k=1}^{K}$
of the input space $\mathcal{X}$, whose sub-regions are assumed to
be homogeneous in the sense that a co-kriging model \eqref{eq:kenedyohagandynamiclink-1}
can be defined independently at each sub-region, i.e. 
\begin{align}
y_{k,t}(x)= & \xi_{k,t-1}(x)y_{k,t-1}(x)+\delta_{k,t}(x)\qquad\text{for}\:x\in\mathcal{X}_{k},\,t=2,...,S\,;\label{eq:adfgdghasdfg}
\end{align}
such that input dependencies are represented accurately enough by
parameterizing the unknown scale discrepancies $\{\xi_{k,t}(x)\}$,
location discrepancies $\{\delta_{k,t}(x)\}$, and output functions
$\{y_{k,1}(x)\}$ with computationally convenient forms.

We cast $\{\mathcal{X}_{k}\}_{k=1}^{K}$ as a binary tree partition
with rectangular sub-regions $\mathcal{X}_{k}:=\mathcal{X}_{k}(\mathcal{T})$,
for $k=1,...,K(\mathcal{T})$, determined by a binary tree $\mathcal{T}$.
This specification adds structure to the model for the sake of computational
convenience, however it can still provide a reasonable approximation
to the reality. Binary treed partitioning has been successfully used
in other problems \citep{Denison98,Chipman98,Gramacy08,pratola2017heteroscedastic,konomi2016_JASA,KaragiannisKonomiLin2017}.
To account for the uncertainty about $\mathcal{T}$, we use the binary
tree process prior of \citet{Chipman98} specified as 
\begin{equation}
\pi(\mathcal{T})=P_{\text{rule}}(\rho|v,\mathcal{T})\prod_{v_{i}\in\mathcal{I}}P_{\text{split}}(v_{i},\mathcal{T})\prod_{v_{j}\in\mathcal{E}}(1-P_{\text{split}}(v_{j},\mathcal{T})),\label{eq:rwtwetr}
\end{equation}
where $\mathcal{E}$ denotes the set of external nodes corresponding
to sub-regions of the partition $\{\mathcal{X}_{k}(\mathcal{T})\}$
and $\mathcal{I}$ denotes the internal nodes. Tree $\mathcal{T}$
has origin denoting the whole input space $\mathcal{X}$, while each
node $v\in\mathcal{T}$ represents a sub-region of the input space.
Each node splits with probability $P_{\text{split}}(v,\mathcal{T})=\zeta(1+u_{v})^{-d}$
where $u_{v}$ is the depth of $v\in\mathcal{\mathcal{T}}$, $\zeta$
controls the balance of the shape of the tree, and $d$ controls the
size of the tree. The splits are preformed based on a random splitting
rule $\rho$ following a distribution $P_{\text{rule}}(\rho|v,\mathcal{T})$.

We specify mutually independent Gaussian processes (GP) priors for
$y_{k,1}(\cdot)$, and $\delta_{k,t}(\cdot)$ 
\begin{align}
y_{k,1}(\cdot) & |\mathcal{T}\sim\text{GP}(\mu_{1}(\cdot|\beta_{k,1}),\sigma_{k,1}^{2}R_{1}(\cdot,\cdot|\phi_{k,1}));\label{eq:etastatgp}\\
\delta_{k,t}(\cdot) & |\mathcal{T}\sim\text{GP}(\mu_{t}(\cdot|\beta_{k,t}),\sigma_{k,t}^{2}R_{t}(\cdot,\cdot|\phi_{k,t})),\:\text{for}\,t=2,\dots,S,\label{eq:deltastatgp}
\end{align}
for $k=1,...,K$, to account for their uncertainty. Given a suitable
partition $\{\mathcal{X}_{k}\}_{k=1}^{K}$ for the model \eqref{eq:adfgdghasdfg},
we can use simple and computationally convenient functions to model
$\mu_{t}(\cdot|\beta_{k,t})$, $R_{t}(\cdot,\cdot|\phi_{k,t})$, and
$\xi_{k,t}(x)$. We specify square exponential correlation function
in separable form $R_{t}(x,x'|\phi_{k,t})=\exp(-\frac{1}{2}(x-x')^{\top}\text{diag}(\phi_{k,t})(x-x'))$,
however more sophisticated ones can be used \citep{williams2006gaussian}.
The mean functions are parametrized as basis expansions $\mu_{t}(\cdot|\beta_{k,t})=h_{t}(\cdot)^{T}\beta_{k,t}$,
where $h_{t}(\cdot)$ is a vector of basis functions and $\beta_{k,t}$
are vectors of coefficients, at fidelity level $t$, and sub-region
$\mathcal{X}_{k}$. The unknown functions $\{\xi_{k,t}(x)\}$ are
modeled as low degree basis expansions $\xi_{k,t}(x|\gamma_{k,t})=w_{t}(x)^{T}\gamma_{k,t}$
where $\{w_{t}(x)\}$ are polynomial bases and $\{\gamma_{k,t}\}$
are uncertain coefficients. Modeling $\mu_{t}(\cdot|\beta_{k,t})$,
and $\xi_{k,t}(x)$ as basis expansions facilitates the specification
of conjugate priors and leads to computational savings given a suitable
treatment in the likelihood.

\subsection{Conditional-conjugacy via augmentation\label{sec:Bayesian-inference-and}}

We do not require the available experimental design to be hierarchically
nested, unlike existing co-kriging methods \citep{KennedyOhagan2000,Gratiet_SIAM_UQ2013}.
Namely, if $\{y_{t},\mathfrak{X}_{t}\}$ denotes the available a training
data-set with output values $y_{t}\in\mathbb{R}^{n_{t}}$ at the experimental
design $\mathfrak{X}_{t}$ of size $n_{t}$ at fidelity level $t=1,..,S$,
it may be $\mathfrak{X}_{t+1}\not\subseteq\mathfrak{X}_{t}$ for some
$t$. This realistic generalization prevents the direct specification
of priors conjugate to the Gaussian likelihood $f(y_{1:S}|\mathcal{T},\sigma_{1:S}^{2},\phi_{1:S},\beta_{1:S},\gamma_{1:S-1})$,
and hence makes the Bayesian computations prohibitively expensive.
In such cases, direct implementation of existing co-kriging methods
would require the inversion of large covariance matrices with size
$\sum_{t}n_{t}\times\sum_{t}n_{t}$ for the computation of the likelihood,
and possibly the use of Metropolis-Hastings operations in high-dimensional
state spaces which would lead to practically infeasible computations.
The introduction of the binary partition aggregates this issue as
it increases the dimensionality of the posterior by introducing additional
unknown parameters $\beta_{k,t},\gamma_{k,t},\sigma_{k,t}^{2},\phi_{k,t}$;
this necessitates the specification of conjugate priors.

We address this issue by properly imputing the observed data with
uncertain quantities, that can be thought of as missing data of a
hierarchically nested experimental design able to induce a conditional
independence that enables the specification of conjugate priors, facilitates
tractability of posterior marginals and conditionals, and allows the
design of efficient MCMC implementations, while it leads to the same
Bayesian inference as if we had considered the original data-set only.

\paragraph*{Augmentation}

Let $\{y_{k,t},\mathfrak{X}_{k,t}\}$ be the observed data-set with
output values $y_{k,t}=y_{t}(\mathfrak{X}_{k,t})$ and design $\mathfrak{X}_{k,t}$
at sub-region $\mathcal{X}_{k}$ and fidelity level $t$. Assume sets
of points $\tilde{\mathfrak{X}}_{k,t}$ and $\mathring{\mathfrak{X}}_{k,t}$
such that $\tilde{\mathfrak{X}}_{k,S}=\mathfrak{X}_{k,S}$ with $\mathring{\mathfrak{X}}_{k,S}=\emptyset$,
and $\tilde{\mathfrak{X}}_{k,t}=\mathfrak{X}_{k,t}\cup\mathfrak{\mathring{X}}_{k,t}$
where $\mathfrak{\mathring{X}}_{k,t}=\mathfrak{\tilde{X}}_{k,t+1}\cap(\mathfrak{X}_{k,t})^{\complement}$
for $t=S-1,...,1$. It is easy to check that $\tilde{\mathfrak{X}}_{k,t}=\cup_{j=t}^{S}\mathfrak{X}_{k,j}$,
and that $\{\tilde{\mathfrak{X}}_{k,t}\}_{t=1}^{S}$ is hierarchically
nested; i.e. $\tilde{\mathfrak{X}}_{k,t}\subseteq\tilde{\mathfrak{X}}_{k,t-1}$.
By construction, $\{\mathring{\mathfrak{X}}_{k,t}\}$ is the smallest
collection of sets of input points required to be added to the original
design $\{\mathfrak{X}_{k,t}\}$ in order to obtain a hierarchically
nested experimental design $\{\tilde{\mathfrak{X}}_{k,t}\}$. Let
$\mathring{y}_{k,t}=y_{t}(\mathring{\mathfrak{X}}_{k,t})$ be the
missing output values of the computer model at the corresponding input
points in $\mathring{\mathfrak{X}}_{k,t}$ . We refer to $\{\mathring{y}_{k,t},\mathfrak{\mathring{X}}_{k,t}\}$
as missing data-set, and $\{\tilde{y}_{k,t},\tilde{\mathfrak{X}}_{k,t}\}$
as complete data-set, where $\tilde{\mathfrak{X}}_{k,t}$ is the complete
experimental design, and $\tilde{y}_{k,t}=y_{t}(\tilde{\mathfrak{X}}_{k,t})$
are the output model values at input points in $\tilde{\mathfrak{X}}_{k,t}$.

The joint distribution of $\tilde{y}=(\tilde{y}_{k,t})$ given the
parameters $(\mathcal{T},\beta,\gamma,\sigma^{2},\phi)$ is 
\begin{align}
f(\tilde{y}|\mathcal{T},\beta,\gamma,\sigma^{2},\phi)=\prod_{k=1}^{K} & f_{k}(\tilde{y}_{k,1}|\beta_{k,1},\sigma_{k,1}^{2},\phi_{k,1})\prod_{t=2}^{S}f_{k}(\tilde{y}_{k,t}|\tilde{y}_{k,t-1},\beta_{k,t},\gamma_{k,t-1},\sigma_{k,t}^{2},\phi_{k,t})\label{eq:likelihood}
\end{align}
where each conditional $f_{k}(\tilde{y}_{k,t}|...)$ is a Gaussian
distribution with mean $\xi{}_{t-1}(\tilde{\mathfrak{X}}_{k,t}|\gamma_{k,t-1})\circ y_{k,t-1}(\tilde{\mathfrak{X}}_{k,t})+\mu_{t}(\tilde{\mathfrak{X}}_{k,t}|\beta_{k,t})$,
and covariance $\sigma_{k,t}^{2}R_{t}(\tilde{\mathfrak{X}}_{k,t},\tilde{\mathfrak{X}}_{k,t}|\phi_{k,t})$.
Here, $\circ$ denotes the Hadamard product. The join distribution
of $\tilde{y}$ can be factorized as in \eqref{eq:likelihood} because
the proposed augmentation artificially creates a hierarchically nested
design which due to the Markovian condition of \eqref{eq:adfgdghasdfg}
induces the required conditional independence. The computation of
the augmented likelihood \eqref{eq:likelihood} is broken down into
that of $S$ Gaussian densities requiring the inversion of $\tilde{n}_{k,t}\times\tilde{n}_{k,t}$
covariance matrices. Otherwise, we would be unable to factorize \eqref{eq:likelihood}
and we would be required to invert a larger covariance matrices with
sizes $\sum_{t}n_{t}\times\sum_{t}n_{t}$.

\paragraph*{Priors}

To account for the uncertainty about unknowns $\beta,\gamma,\sigma^{2},\phi$,
we specify a prior factorized as 
\begin{equation}
\pi(\beta,\gamma,\sigma^{2},\phi|\mathcal{T})=\prod_{k=1}^{K}\pi(\beta_{k,1},\sigma_{k,1}^{2}|\mathcal{T})\pi(\phi_{k,1}|\mathcal{T})\prod_{t=2}^{S}\pi(\beta_{k,t},\gamma_{k,t-1},\sigma_{k,t}^{2}|\mathcal{T})\pi(\phi_{k,t}|\mathcal{T}).\label{eq:sdgsg}
\end{equation}

We assign Normal-inverse-gamma prior distributions on $(\beta,\gamma,\sigma^{2})$
such as 
\begin{align*}
\beta_{k,1}|\mathcal{T},\sigma_{k,1}^{2} & \sim\text{N}_{p_{1}}(b_{1},\sigma_{k,1}^{2}B_{1})\,; & \sigma_{k,1}^{2}|\mathcal{T} & \sim\text{IG}(\lambda_{1},\chi_{1})\,;\\
\beta_{k,t},\gamma_{k,t-1}|\mathcal{T},\sigma_{k,t}^{2} & \sim\text{N}_{p_{t}+q_{t-1}}(\left[b_{t},g_{t-1}\right]^{\top},\sigma_{k,t}^{2}\text{diag}\left(B_{t},G_{t-1}\right)^{\top})\,; & \sigma_{k,t}^{2}|\mathcal{T} & \sim\text{IG}(\lambda_{t},\chi_{t})\,;
\end{align*}
which are conjugate to the conditionals $f_{k}(\tilde{y}_{k,t}|...)$
in augmented likelihood \eqref{eq:likelihood}. This allows the analytic
marginalization of the posterior and leads to important computational
benefits discussed in Section \ref{subsec:Bayesian-inference-and}.
Without augmentation, we would be unable to specify conjugate priors
for the actual likelihood, and computations for learning $(\beta,\gamma,\sigma^{2})$
would be impractical. Elicitation of the priors is performed according
to \citep{Oakley2002,brynjarsdottir2014learning}. Weakly informative
Jeffreys' priors are obtained by adjusting $b_{t}$, $g_{t-1}$, $B_{t}^{-1}$
and $G_{t}^{-1}$ to be close to zero, and $\lambda_{t}\rightarrow1+(p_{t}+q_{t-1})/2$
for $t=2,...,S$, and $\lambda_{1}\rightarrow1+p_{1}/2$. Here, $\{\pi(\text{\ensuremath{\phi_{k,t}}}|\mathcal{T})\}$
are proper priors chosen by the researcher.

The posterior distribution of ABTCK model is 
\begin{equation}
\pi(\mathcal{T},\beta,\gamma,\sigma^{2},\phi,\mathring{y}|y)\propto f(\mathring{y}|y,\mathcal{T},\beta,\gamma,\sigma^{2},\phi)f(y|\mathcal{T},\beta,\gamma,\sigma^{2},\phi)\pi(\beta,\gamma,\sigma^{2},\phi|\mathcal{T})\pi(\mathcal{T}),\label{eq:adfhgdfhzsdgh}
\end{equation}
admits the posterior of interest $\pi(\mathcal{T},\beta,\gamma,\sigma^{2},\phi|y)$
as marginal by construction, and hence leads to the same Bayesian
analysis.

\subsection{Bayesian inference and computations\label{subsec:Bayesian-inference-and}}

We design a RJMCMC sampler, targeting the augmented posterior \eqref{eq:adfhgdfhzsdgh},
that involves a random permutation scan of blocks updating $[\mathring{y}|y,\phi,\sigma^{2},\gamma,\mathcal{T}]$,
$[\phi,\mathcal{T}|\tilde{y}]$, and $[\beta,\gamma,\sigma^{2},\phi|\tilde{y},\mathcal{T}]$.
The blocks are collapsed to avoid undesired high MC standard errors
due to the originally high-dimensional sampling space \citep{liu1994collapsed}.
The sampler is computationally efficient as it breaks down the inversion
of covariance matrices and involves parallel sampling at different
sub-regions $k$ and fidelity levels $t$. Details regarding the MCMC
blocks are explained below.

\paragraph*{Update $[\mathring{y}|y,\phi,\gamma,\sigma^{2},\mathcal{T}]$ \label{sec:Pro2}}

The full conditional posterior of $\mathring{y}_{k,t}$, after integrating
out $\beta$'s from the joint posterior \eqref{eq:adfhgdfhzsdgh},
is a Normal distribution with mean and covariance matrix 
\begin{align}
\mathring{\mu}_{k,t}= & \mathring{\Sigma}_{k,t}\left[\sigma_{k,t}^{-2}\hat{R}_{t}^{-1}(\phi_{k,t}|\mathring{\mathfrak{X}}_{k,t};\mathfrak{X}_{k,t})\hat{\mu}_{(t-1)\rightarrow t}(\phi_{k,t},\gamma_{k,t-1}|\mathring{\mathfrak{X}}_{k,t};\mathfrak{X}_{k,t})+\Xi_{t}(\mathring{\mathfrak{X}}_{k,t}|\gamma_{k,t})\right.\nonumber \\
 & \times\left.\sigma_{k,t+1}^{-2}\hat{R}_{t+1}^{-1}(\phi_{k,t+1}|\mathring{\mathfrak{X}}_{k,t};\tilde{\mathfrak{X}}_{k,t+1}\cap\mathring{\mathfrak{X}}_{k,t}^{\complement})\hat{\mu}_{(t+1)\rightarrow t}(\phi_{k,t+1},\gamma_{k,t}|\mathring{\mathfrak{X}}_{k,t};\tilde{\mathfrak{X}}_{k,t+1}\cap\mathring{\mathfrak{X}}_{k,t}^{\complement})\right]\label{eq:sdjksdbfg}\\
\mathring{\Sigma}_{k,t}= & \left[\frac{\hat{R}_{t}^{-1}(\phi_{k,t}|\mathring{\mathfrak{X}}_{k,t};\mathfrak{X}_{k,t})}{\sigma_{k,t}^{2}}+\Xi_{t}(\mathring{\mathfrak{X}}_{k,t}|\gamma_{k,t})\frac{\hat{R}_{t+1}^{-1}(\phi_{k,t+1}|\mathring{\mathfrak{X}}_{k,t};\tilde{\mathfrak{X}}_{k,t+1}\cap\mathring{\mathfrak{X}}_{k,t}^{\complement})}{\sigma_{k,t+1}^{2}}\Xi_{t}(\mathring{\mathfrak{X}}_{k,t}|\gamma_{k,t})\right]^{-1}\nonumber 
\end{align}
where $\Xi_{t}(\mathring{\mathfrak{X}}_{k,t}|\gamma_{k,t})=\text{diag}(\xi_{t}(\mathring{\mathfrak{X}}_{k,t}|\gamma_{k,t}))$,
for $k=1,...,K$ and $t=1,...,S-1$. The functions $\hat{R}_{t}$,
$\hat{\mu}_{(t-1)\rightarrow t}$, and $\hat{\mu}_{(t+1)\rightarrow t}$
are given in the Appendix \ref{sec:Appendix}. We observe that, updating
missing data $\mathring{y}_{k,t}$ takes into account information
from the lower level $t-1$, the current level $t$, and higher level
$t+1$ by interpolating the associated moments. For instance, $\hat{\mu}_{(t-1)\rightarrow t}$
(and $\hat{\mu}_{(t+1)\rightarrow t}$) provide information about
the location of $\mathring{y}_{k,t}$ from levels $t-1$, $t$ (and
levels $t+1$, $t$). Hence, each update interpolates not only across
the input space at an individual fidelity level but also across the
fidelity levels. Simulation of $[\mathring{y}_{k,t}|y,\phi,\gamma,\sigma^{2},\mathcal{T}]$
can be performed in parallel for $k$ which is a computational benefit,
and it can be suppressed if $\mathring{\mathfrak{X}}_{k,t}=\emptyset$.

Elaborating further into specific cases of the above imputation, if
levels $t$ and $t+1$ do not share any design points at all, at sub-region
$\mathcal{X}_{k}$, i.e., $\tilde{\mathfrak{X}}_{k,t+1}\cap\mathring{\mathfrak{X}}_{k,t}^{\complement}=\emptyset$,
then $\hat{R}_{t+1}^{-1}(\phi_{k,t+1}|\mathring{\mathfrak{X}}_{k,t};\emptyset)=R_{t+1}^{-1}(\mathring{\mathfrak{X}}_{k,t},\mathring{\mathfrak{X}}_{k,t}|\phi_{k,t+1})$,
and $\hat{\mu}_{(t+1)\rightarrow t}(\phi_{k,t+1},\gamma_{k,t}|\mathring{\mathfrak{X}}_{k,t};\emptyset)=y_{k,t+1}(\mathring{\mathfrak{X}}_{k,t})-H_{t+1}(\mathring{\mathfrak{X}}_{k,t})b_{t+1}$.
This implies that, given weak priors on $\delta_{k,t+1}(\cdot)$ are
specified, i.e. $b_{t+1}\rightarrow0$, the update of missing $\mathring{y}_{k,t}$
obtains information from the upper level $t+1$ which entirely relies
on the observed output $y_{k,t+1}$ and not from the discrepancy terms
$\delta_{k,t+1}(\cdot)$ and $\xi_{k,t}(\cdot)$ of the two levels.
If levels $t$ and $t+1$ share design points, $\tilde{\mathfrak{X}}_{k,t+1}\cap\mathring{\mathfrak{X}}_{k,t}^{\complement}\ne\emptyset$,
the extra structure of the equations of $\hat{\mu}_{(t+1)\rightarrow t}$
and $\hat{R}_{t+1}^{-1}$ in \eqref{eq:wtytywet} and \eqref{eq:etywsywet}
(see Appendix \ref{sec:Appendix}) can be interpreted as the factor
quantifying the discrepancy between levels $t$ and $t+1$. Finally,
we can see that when the correlation between the two levels $t$ and
$t+1$, at sub-region $\mathcal{X}_{k}$, is weak, e.g. $\Xi_{t}(\mathring{\mathfrak{X}}_{k,t}|\gamma_{k,t})\rightarrow0$,
the missing data update resembles the prediction relying only on the
information from the current level $t$. Based on these observations,
it may be preferable to consider designs with some overlap at adjacent
levels not only for computational convenience but also for modeling
reasons. However, a theoretical proof of this statement is out of
our scope.

\paragraph*{Update $[\mathcal{T},\phi|\tilde{y}]$ \label{sec: Bayesian inf2}}

To update $[\mathcal{T},\phi|\tilde{y}]$, we propose a mixture of
the Markov transitions targeting the augmented marginal posterior
$\pi(\mathcal{T},\phi|\tilde{y})$ whose density is proportional to

\begin{align}
\pi(\tilde{y},\mathcal{T},\phi) & =\pi(\mathcal{T})\prod_{k=1}^{K}\pi(\tilde{y}_{k,1},\phi_{k,1}|\mathcal{T})\prod_{t=2}^{S}\pi(\tilde{y}_{k,t},\phi_{k,t}|\tilde{y}_{k,t-1},\mathcal{T}),\label{eq:phiposterior-1}\\
\pi(\tilde{y}_{k,t},\phi_{k,t}|\tilde{y}_{k,t-1},\mathcal{T}) & =\pi(\phi_{k,t})\frac{|\hat{A}_{k,t}(\phi_{k,t})|^{\frac{1}{2}}}{|B_{t}|^{\frac{1}{2}}|G_{t}|^{\frac{1}{2}}}\frac{\chi_{t}^{\lambda_{t}}}{\pi^{\frac{\tilde{n}_{k,t}}{2}}}\frac{\Gamma(\lambda_{t}+\frac{\tilde{n}_{k,t}}{2})}{\Gamma(\lambda_{t})}\left(\text{SSE}_{k,t}(\phi_{k,t})\right)^{-\lambda_{t}-\frac{\tilde{n}_{k,t}}{2}}\label{eq:marginal likelihoodwithphi-1}
\end{align}
where $\text{SSE}_{k,t}(\phi_{k,t})=(\tilde{n}_{k,t}+2\lambda_{t}-2)\hat{\sigma}_{k,t}^{2}(\phi_{k,t})$.
Functions $\hat{\sigma}_{k,t}^{2}$ and $\hat{A}_{k,t}$ are given
in \eqref{eq:dghdfghdafsg} and \eqref{eq:wthtg} in Appendix \ref{sec:Appendix}.
The Markov transitions are based on the operations change, swap, rotate,
and grow \& prune, introduced by \citep{Chipman98,Gramacy08}. The
first three operations are Metropolis-Hastings algorithms \citep{Hastings1970}
whose implementation is straightforward. The grow \& prune operations
are local reversible jump (RJ) transitions and further specification
is required.

The grow operation performing a transition from state $(\mathcal{T},\phi)$
to $(\mathcal{T}^{*},\phi^{*})$ works as follows. We randomly select
an external node $\omega_{j_{0}}$ and assume it corresponds to a
sub-region $\mathcal{X}_{j_{0}}$, data-set $\{\mathfrak{\tilde{X}}_{j_{0}},\tilde{y}_{j_{0}}\}$,
and parameters $\phi_{j_{0},t}$ though the augmented statistical
model. We propose node $\omega_{j_{0}}$ to split into two new child
nodes $\omega_{j_{1}}$ and $\omega_{j_{2}}$ according to the splitting
rule $P_{\text{rule}}$ in prior \eqref{eq:sdgsg}, and we denote
the proposed tree as $\mathcal{T}^{*}$. Nodes $\omega_{j_{1}}$ and
$\omega_{j_{2}}$ correspond to disjoint sub-regions $\mathcal{X}_{j_{1}}$
and $\mathcal{X}_{j_{2}}$ (with $\mathcal{X}_{j_{0}}=\mathcal{X}_{j_{0}}\cup\mathcal{X}_{j_{1}}$),
data-sets $\{\mathfrak{\tilde{X}}_{j_{1},t},\tilde{y}_{j_{1},t}\}$
and $\{\mathfrak{\tilde{X}}_{j_{2},t},\tilde{y}_{j_{2},t}\}$, and
parameters $\phi_{j_{1},t}^{*}$ and $\phi_{j_{2},t}^{*}$, respectively.
Randomly, one of the parameters $\phi_{j_{1},t}^{*}$ or $\phi_{j_{2},t}^{*}$
inherits the values from the parent ones; e.g., $\phi_{j_{1},t}^{*}=\phi_{j_{0},t}$.
The values of the other parameter are proposed by simulating from
a probability distribution; e.g., $\phi_{j_{2},t}^{*}\sim Q_{t}(\cdot)$,
such as the corresponding priors. The rest elements of $\phi_{t}^{*}$
inherit their values from $\phi_{t}$. The proposed transition is
accepted with probability $\min(1,A)$ where 
\begin{align}
A= & \frac{\zeta(1+u_{\omega_{j_{0}}})^{-d}(1-\zeta(2+u_{\omega_{j_{0}}})^{-d})^{2}}{1-\zeta(1+u_{\omega_{j_{0}}})^{-d}}\frac{|\mathcal{G}|}{|\mathcal{P}^{*}|}\prod_{t=2}^{S}\frac{\pi(\tilde{y}_{j_{1},t},\phi_{j_{1}}^{*}|\tilde{y}_{j_{1},t-1},\mathcal{T}^{*})\pi(\tilde{y}_{j_{2},t},\phi_{j_{2},t}^{*}|\tilde{y}_{j_{2},t-1},\mathcal{T}^{*})}{\pi(\tilde{y}_{j_{0},t},\phi_{j_{1},t}^{*}|\tilde{y}_{j_{0},t-1},\mathcal{T})Q_{t}(\phi_{j_{2},t}^{*})}\nonumber \\
 & \qquad\qquad\times\frac{\pi(\tilde{y}_{j_{1},1},\phi_{j_{1},1}^{*}|\mathcal{T}^{*})\pi(\tilde{y}_{j_{2},1},\phi_{j_{2},1}^{*}|\mathcal{T}^{*})}{\pi(\tilde{y}_{j_{0},1},\phi_{j_{1},1}^{*}|\mathcal{T})Q_{t}(\phi_{j_{2},1}^{*})},\label{eq:RJACCRATIO}
\end{align}
$\mathcal{G}$ is the set of growable nodes in tree $\mathcal{T}$,
and $\mathcal{P}^{*}$ is the set of prounable nodes in tree $\mathcal{T}^{*}$.
The prune operation, performing a transition from state $(\mathcal{T}^{*},\phi^{*})$
to $(\mathcal{T},\phi)$, is fully defined as the reverse operation
of the Grow one, and is accepted with probability $\min(1,1/A)$.

Due to the proposed augmentation in Section \ref{sec:Bayesian-inference-and},
we are able to analytically integrate out a potentially high-dimensional
parameter vector $(\beta,\gamma,\sigma^{2})$ from the joint density
\eqref{eq:adfhgdfhzsdgh}, and hence design local RJ moves targeting
the marginal $\pi(\mathcal{T},\phi|\tilde{y})$. The benefit from
this collapsed update is that the proposed RJ algorithm operates on
a lower dimensional state space, which allows for shorter and more
acceptable jumps in practice. If necessary, grow and prune operations
can be further improved by using the annealing mechanism of \citet{KaragiannisAndrieu2013}.

\paragraph*{Update $[\beta,\gamma,\sigma^{2},\phi|\tilde{y},\mathcal{T}]$}

The conditional posterior $\pi(\beta,\gamma,\sigma^{2},\phi|\tilde{y},\mathcal{T})$
has the form 
\begin{align}
\beta_{k,t}|\tilde{y}_{k,t},\tilde{y}_{k,t-1},\gamma_{k,t-1},\sigma_{k,t}^{2},\phi_{k,t}\sim & \text{N}(\hat{\beta}_{k,t}(\phi_{k,t}),\hat{B}_{k,t}(\phi_{k,t})\sigma_{k,t}^{2}),\,\text{for }t=2,...S\label{eq:pouitsesmple}\\
\beta_{k,1}|\tilde{y}_{k,1},\sigma_{k,1}^{2},\phi_{k,1}\sim & \text{N}(\hat{\beta}_{k,1}(\phi_{k,1}),\hat{B}_{k,1}(\phi_{k,1})\sigma_{k,1}^{2}),\label{eq:olympiakara}\\
\gamma_{k,t-1}|\tilde{y}_{k,t},\tilde{y}_{k,t-1},\sigma_{k,t}^{2},\phi_{k,t}\sim & \text{N}(\hat{\gamma}_{k,t-1}(\phi_{k,t}),\hat{G}_{k,t-1}(\phi_{k,t})\sigma_{k,t}^{2}),\,\text{for }t=2,...S\nonumber \\
\sigma_{k,t}^{2}|\tilde{y}_{k,t},\tilde{y}_{k,t-1},\phi_{k,t}\sim & \text{IG}(\hat{\lambda}_{k,t},\hat{\chi}_{k,t}(\phi_{k,t})),\,\text{for }t=2,...S\label{eq:thrilole}\\
\sigma_{k,1}^{2}|\tilde{y}_{k,1},\phi_{k,1}\sim & \text{IG}(\hat{\lambda}_{k,1},\hat{\chi}_{k,1}(\phi_{k,1})),\label{eq:antegeiametakitsoukala}\\
\phi_{k,t}|\tilde{y},\mathcal{T}\sim & \text{d}\pi(\phi_{k,t}|\tilde{y},\mathcal{T}),\label{eq:asdfgsdf}
\end{align}
where the hatted quantities are given in \eqref{eq:trfjhfghdfh}-\eqref{eq:dghdfghdafsg}
of Appendix \ref{sec:Appendix}.

Conditional distributions \eqref{eq:pouitsesmple}-\eqref{eq:antegeiametakitsoukala}
can be sampled directly, and in parallel for different $(k,t)$. Sampling
from the full conditional of $\beta$'s \eqref{eq:pouitsesmple} and
\eqref{eq:olympiakara} is not necessary and can be ignored from the
MCMC swap if prediction is the only concern of the analysis. This
is because $\beta$'s can be analytically integrated out from the
proposed emulator in Section \ref{subsec:Posterior-analysis-and}.
Alternatively, $\beta$'s can be sampled outside the MCMC swap \eqref{eq:pouitsesmple}
and \eqref{eq:olympiakara} by conditioning.

Updating $\phi$ by simulating from $\pi(\phi|\tilde{y},\mathcal{T})$
is not necessary in theory, as it is updated in block $[\mathcal{T},\phi|\tilde{y}]$,
however it improves mixing in practice. The marginal posterior \eqref{eq:asdfgsdf}
cannot be sampled directly. Conditional independence in \eqref{eq:phiposterior-1}
implies that $\{\phi_{k,t}\}$ can be simulated by running in parallel
$K\times S$ Metropolis-Hastings algorithms each of them targeting
distributions with densities proportional to \eqref{eq:marginal likelihoodwithphi-1}.

\subsection{Posterior analysis and emulation \label{subsec:Posterior-analysis-and}}

Assume there is available a MCMC sample $\mathcal{S}^{N}=(\mathring{y}^{(j)},\mathcal{T}^{(j)},\gamma^{(j)},\sigma^{2,(j)},\phi^{(j)})_{j=1}^{N}$
generated from the RJMCMC sampler in Section \ref{subsec:Bayesian-inference-and},
and let $\{\mathcal{X}_{k}^{(j)}\}_{k=1}^{K^{(j)}}$ denote the partition
corresponding to tree $\mathcal{T}^{(j)}$. Central Limit Theorem
can be applied to facilitate inference as the proposed sampler is
aperiodic, irreducible, and reversible \citep{roberts2004general}.

The proposed procedure ABTCK allows inference to be performed for
the missing output values $\mathring{y}_{t}=y_{t}(\mathring{\mathfrak{X}}_{t})$
at input points in $\mathring{\mathfrak{X}}_{t}=\bigcup_{\forall k}\mathring{\mathfrak{X}}_{k,t}$.
Inference on $\mathring{y}_{t}$ can be particularly useful when the
computer model has been unable to generate simulations at these input
points due to numerical crash or limitations. The marginal posterior
distribution of $\mathring{y}_{t}$, along with its expectations,
can be approximated via standard Monte Carlo (MC) using the generated
samples $\{\mathring{y}_{t}^{(j)}\}$ at each level $t$. Alternatively,
point estimates of $\mathring{y}_{k,t}$ at $\mathring{\mathfrak{X}}_{k,t}$
can be approximated by the more accurate Rao-Blackwell MC estimator
$\text{E}(\mathring{y}_{k,t}|y_{1:S})\approx\frac{1}{N}\sum_{j=1}^{N}\mathring{\mu}_{k,t}^{(j)},$
where $\{\mathring{\mu}_{k,t}^{(j)}\}$ is the $j$-th MCMC realization
of \eqref{eq:sdjksdbfg}.

A Monte Carlo recursive emulator able to facilitate fully Bayesian
predictive inference on the output $y_{t}(\mathfrak{X}^{*})$ at untried
input points $\mathfrak{X}^{*}$ at every fidelity level $t=1,...,S$
can be derived. The conditional distribution $[y_{1:S}(\cdot)|y_{1:S},\mathring{y}_{1:S},\beta_{1:S},\gamma_{1:S},\sigma_{1:S}^{2},\phi_{1:S}]$
inherits a conditional independence similar to \eqref{eq:likelihood}
due to the augmentation of the data with $\mathring{y}_{1:S}$ that
allows to be analytically integrated out with respect to \eqref{eq:pouitsesmple}-\eqref{eq:antegeiametakitsoukala}.
Hence the distribution of $[y_{1:S}(\cdot)|y_{k,1:S},\mathring{y}_{k,1:S},\phi_{k,1:S},\mathcal{T}]$,
at sub-region $\mathcal{X}_{k}$, is calculated as 
\begin{align}
y_{1}(\cdot)|\mathring{y}_{1},\phi_{1},\mathcal{T}\sim\text{STP}\left(\mu_{k,1}^{*}(\cdot|\mathring{y}_{k,1},\phi_{k,1}),\right.\hat{\sigma}_{k,1}^{2} & \left.R_{k,1}^{*}(\cdot,\cdot|\mathring{y}_{k,1},\phi_{k,1}),2\lambda_{1}+\tilde{n}_{k,1}\right);\label{eq:adhsdghh}\\
y_{t}(\cdot)|y_{t-1}(\cdot),\mathring{y}_{t:t-1},\phi_{k,t},\mathcal{T}\sim\text{STP}\left(\mu_{k,t}^{*}(\cdot|\mathring{y}_{k,t},\phi_{k,t}),\right. & \left.\hat{\sigma}_{k,t}^{2}R_{k,t}^{*}(\cdot,\cdot|\mathring{y}_{k,t},\phi_{k,t}),2\lambda_{t}+\tilde{n}_{k,t}\right),\label{eq:sghsfghf}
\end{align}
where the conditionals are Student-T processes (STP) with 
\begin{align*}
\mu_{t}^{*}(x|\mathring{y}_{k,t},\phi_{k,t})= & L_{t}(x;y_{t})\hat{a}_{t}+R_{t}(x,\tilde{\mathfrak{X}}_{t}|\phi_{k,t})R_{t}^{-1}(\tilde{\mathfrak{X}}_{k,t},\tilde{\mathfrak{X}}_{k,t}|\phi_{k,t})\left[L_{t}(\tilde{\mathfrak{X}}_{k,t};y_{t})\hat{a}_{t}-y_{t}(\tilde{\mathfrak{X}}_{k,t})\right]\\
R_{t}^{*}(x,x'|\phi_{k,t})= & R_{t}(x,x'|\phi_{k,t})-R_{t}(x,\tilde{\mathfrak{X}}_{k,t}|\phi_{k,t})R_{t}^{-1}(\tilde{\mathfrak{X}}_{k,t},\tilde{\mathfrak{X}}_{k,t}|\phi_{k,t})R_{t}^{\top}(x',\tilde{\mathfrak{X}}_{k,t}|\phi_{k,t})\\
 & \qquad+\left[L_{t}(x;y_{t})-R_{t}(x,\tilde{\mathfrak{X}}_{k,t}|\phi_{k,t})R_{t}^{-1}(\tilde{\mathfrak{X}}_{k,t},\tilde{\mathfrak{X}}_{k,t}|\phi_{k,t})L_{t}(\tilde{\mathfrak{X}}_{k,t};y_{t})\right]\hat{A}_{t}\\
 & \qquad\qquad\times\left[L_{t}(x';y_{t})-R_{t}(x',\tilde{\mathfrak{X}}_{k,t}|\phi_{k,t})R_{t}^{-1}(\tilde{\mathfrak{X}}_{k,t},\tilde{\mathfrak{X}}_{k,t}|\phi_{k,t})L_{t}(\tilde{\mathfrak{X}}_{k,t};y_{t})\right]^{\top}
\end{align*}
for $x,x'\in\mathcal{\mathcal{X}}_{k}$, and $L_{t}(\mathfrak{Z};y_{t-1})=\left[H_{t}(\mathfrak{Z}),\text{diag}(y_{t-1}(\mathfrak{Z})W_{t-1}(\mathfrak{Z}))\right]$
for $t=2:S$ and $L_{1}(\mathfrak{Z};\cdot)=H_{1}(\mathfrak{Z})$
for a set $\mathfrak{Z}$. An MCMC sample from the predictive distribution
of $[y_{1:S}(\cdot)|y_{1:S}]$, at $x\in\mathfrak{X}^{*},$ can be
obtained by simulating \eqref{eq:adhsdghh}-\eqref{eq:sghsfghf} given
the sample values $\mathcal{S}^{N}=\{\mathring{y}^{(j)},\phi^{(j)},\mathcal{T}^{(j)}\}$.
This allows the computation of a Monte Carlo approximation of the
emulator of $[y_{t}(\cdot)|y_{1:S}${]}, and its moments, at any fidelity
level $t$. The conditional independence in the predictive distribution
\eqref{eq:adhsdghh} and \eqref{eq:sghsfghf} results because of our
imputation strategy.

The proposed emulator accounts for non-stationariy and discontinuity
by aggregating simpler GP emulators in a Bayesian model averaging
manner, while it integrates uncertainty regarding the unknown `missing
data' $\mathring{y}$ and parameters. It is computationally preferable
compared to existing co-kriging one \citep{KennedyOhagan2000,Gratiet_SIAM_UQ2013}
because it allows the parallel inversion of smaller covariance matrices
with sizes $\tilde{n}_{t,k}\times\tilde{n}_{t,k}$ while the others
require the inversion of a large co-variance matrix of size $\sum_{t=1}^{S}\tilde{n}_{t}\times\sum_{t=1}^{S}\tilde{n}_{t}$.
Moreover, it is able to recover the whole predictive distribution
and its moments, unlike the derivation in \citet{Gratiet_Garnier2014}
where only the predictive mean and variance are derived recursively.
More importantly, it is able to be applied in problems where the training
data set is not hierarchically nested, while its competitors cannot.

\subsection{Further particulars\label{subsec:Further-particulars}}

Two novel co-kriging procedures can be distinguished as special cases
of the proposed ABTCK. In applications where the design is non hierarchically
nested, but the computer model outputs can be assumed as stationary,
one can consider to drop the partitioning by setting $K=1$ and suppressing
the MCMC update $[\mathcal{T},\phi|\tilde{y}]$. We will refer to
this reduced version of ABTCK as \textit{Augmented Bayesian co-kriging
(ABCK).} Unlike standard co-kriging, our ABCK can be applied with
non-nested designs as it makes the computations for training the Bayesian
model or computing the emulator practically feasible. In fact, the
proposed augmentation strategy separates the posterior into conditionally
independent quantities and allows closed form inference for the majority
of the hyper-parameters. Another special case is where the design
is hierarchically nested but the model outputs present non-stationarity,
the imputation mechanism can be dropped by setting $\{\mathring{\mathfrak{X}}_{k,t}\equiv\emptyset\}$
and suppressing the update $[\mathring{y}|y,\phi,\gamma,\sigma^{2},\mathcal{T}]$.
We will refer to this reduced version of ABTCK as \textit{Bayesian
treed co-kriging (BTCK)}. In such a case, BTCK can be preferable to
the standard co-kriging as it can model the aforesaid stationarity
by properly combining simple stationary GPs.

The computational complexity of the proposed ABTCK compared to existing
co-kriging methods is reduced in two ways: a) by breaking the emulation
into $K$ parts via the partitioning, and b) by breaking the emulation
into $S$ parts via the recursively prediction procedure. In ABTCK
the computational complexity of evaluating the augmented likelihood
or the predictive distribution is $\mathcal{O}(\sum_{t=1}^{S}\sum_{k=1}^{K}\tilde{n}_{k,t}^{3})$
in sequential computing environments, while it can be further reduced
to $\mathcal{O}(\sum_{t=1}^{S}\max_{k=1,\dots,K}(\tilde{n}_{k,t})^{3})$
in parallel computing environments since operations at each $k$ can
be performed in parallel. Under non-hierarchical designs, our ABCK
(assuming the partitioning is dropped) requires $\mathcal{O}(\sum_{t=1}^{S}\tilde{n}_{t}^{3})$
for the evaluation of the augmented likelihood or the Monte Carlo
emulator which is smaller than $\mathcal{O}((\sum_{t=1}^{S}n_{t})^{3})$
required by \citep{KennedyOhagan2000,Gratiet_al2014SIAM_UQ} for the
evaluation of the associated likelihoods since $\tilde{n}_{t}\leq n_{t}$.

\section{Case study \label{sec:simulations}}

\begin{figure}[htb!]
\centering \subfloat[Low-level computer model $\mathscr{C}_{1}$\label{fig:Low-level-computer-model}]{\includegraphics[width=0.45\textwidth]{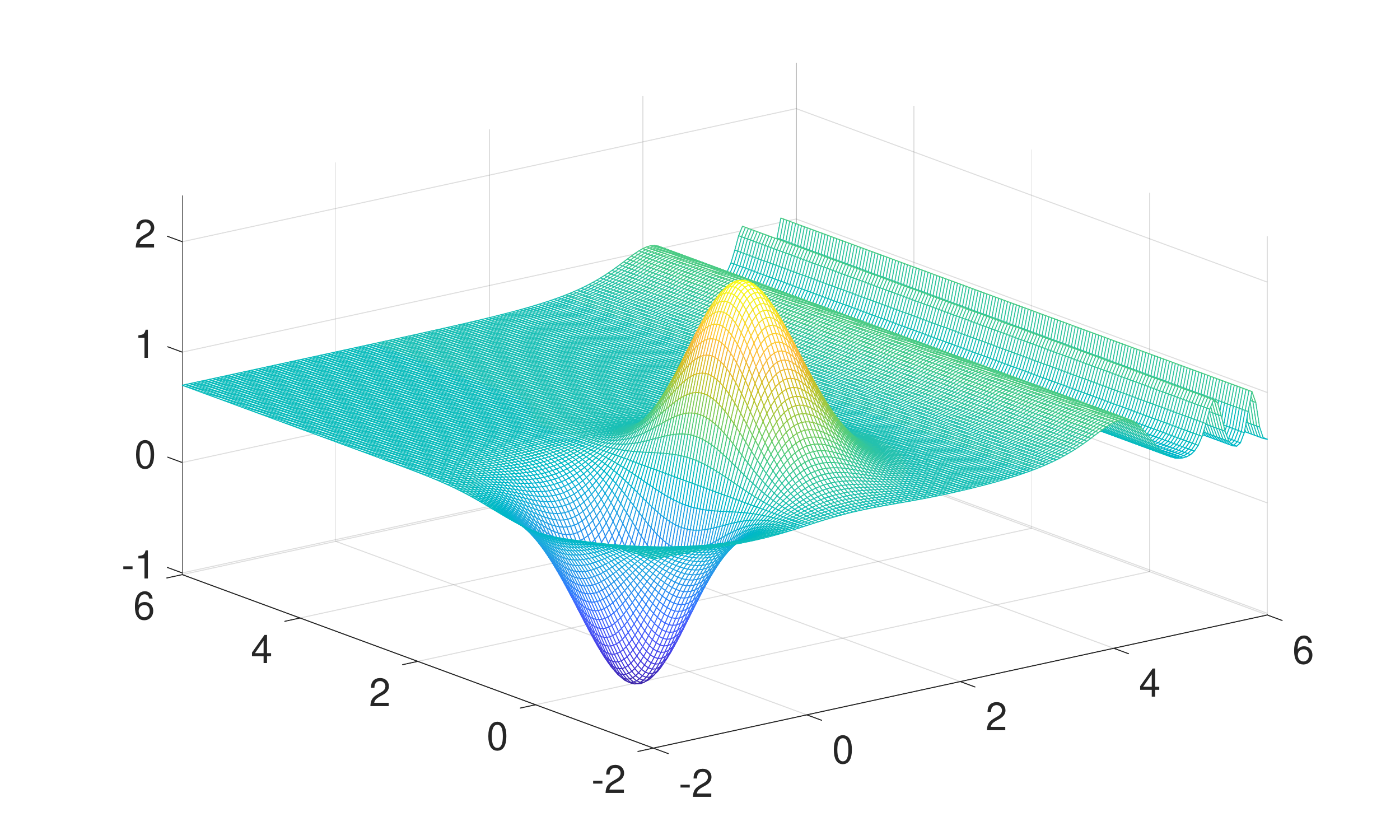}

}\subfloat[High-level computer model $\mathscr{C}_{2}$ \label{fig:High-level-computer-model}]{\includegraphics[width=0.45\textwidth]{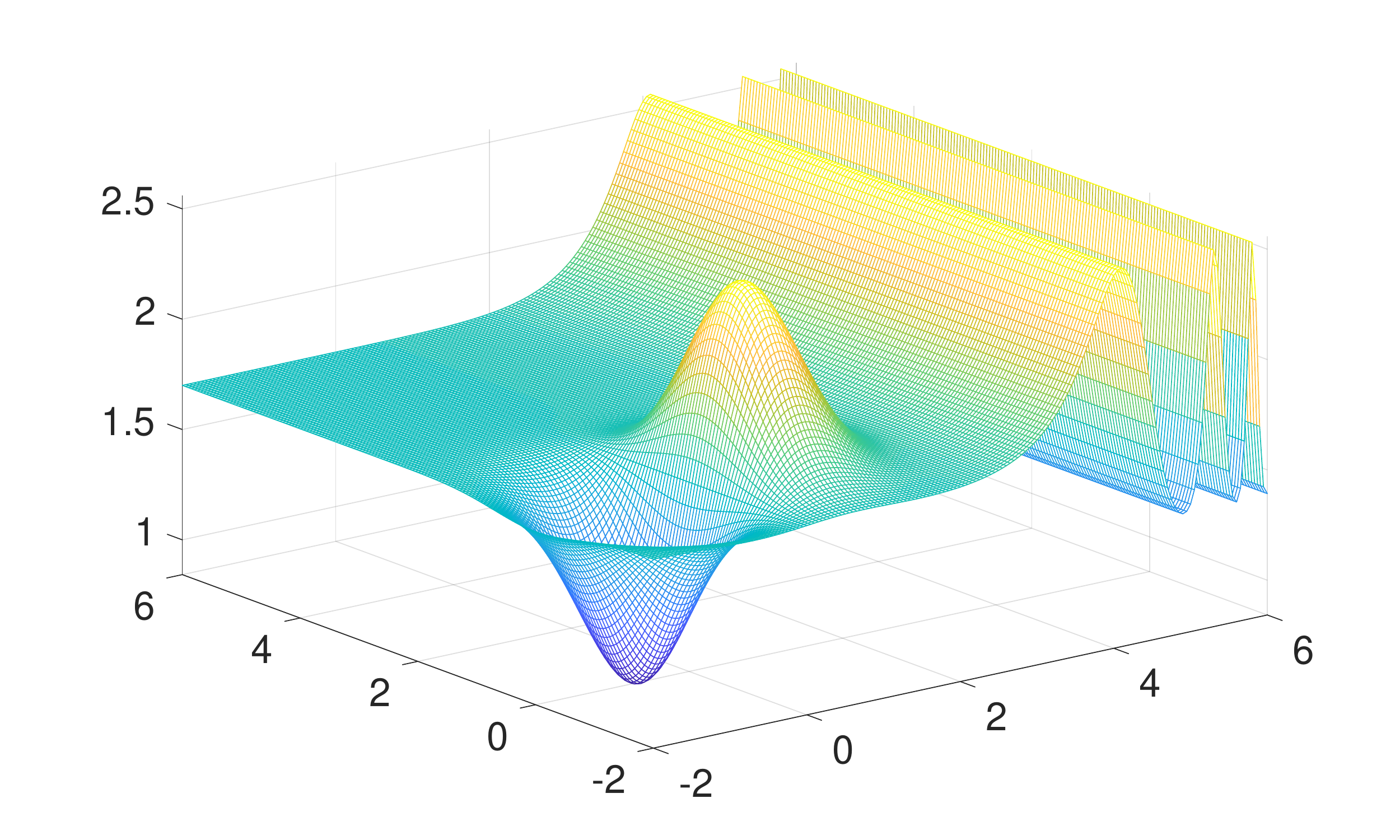}

}

\caption{Exact output functions of the computer model at different fidelity
levels. \label{fig:Real1}}
\end{figure}

We examine the performance of the proposed \textit{augmented Bayesian
treed co-kriging (ABTCK)} method as well as its special case \textit{ABCK
}on a benchmark example. Consider functions 
\begin{equation}
\begin{array}{c}
y_{1}(x)=2x_{1}\exp(-x_{1}^{2}-x_{2}^{2})+0.5\exp\{\sin((0.9(\frac{x_{1}+2}{8}+0.48)^{10}))\}+1.2,\:x\in[-2,6]^{2};\\
y_{2}(x)=4x_{1}\exp(-x_{1}^{2}-x_{2}^{2})+0.2\exp\{\sin((0.9(\frac{x_{1}+2}{8}+0.48)^{10}))\}+0.5,\:x\in[-2,6]^{2},
\end{array}\label{eq:sdfgsdg}
\end{equation}
presented in Figure \ref{fig:Real1}, which are assumed to be output
functions of computer models $\mathscr{C}_{1}$ and $\mathscr{C}_{2}$
with $\mathscr{C}_{2}$ being more accurate but slower to run than
$\mathscr{C}_{1}$. By expressing \eqref{eq:sdfgsdg} as \eqref{eq:kenedyohagandynamiclink-1},
it can be seen that the discrepancy functions $\delta_{1}(\cdot)$
and $\xi_{1}(x)$ change over $\mathcal{X}$. 

We pretend that the equations \eqref{eq:sdfgsdg} are unknown, and
we generate the observed data-set based on a randomly selected non-hierarchically
nested design $\{\mathfrak{X}_{t}\}$. For level $t=1$, the observed
data are generated by employing a Latin Hypercube Sampling (LHS) \citep{McKay1979}
to specify $n_{1}=120$ design points $\mathfrak{X}_{1}$, and computing
the corresponding observations $y_{1}$ from \eqref{eq:sdfgsdg}.
For level $t=2$, the observations are generated likewise by specifying
$n_{2}=30$ input values via LHS such that $\mathfrak{X}_{2}\not\subseteq\mathfrak{X}_{1}$.

We study the effectiveness of the treed partition mechanism in the
co-kriging setting by comparing two versions of the proposed method,
the ABTCK equipped with a partitioning mechanism and the ABCK where
that mechanism is suppressed. Existing co-kriging methods in \citep{KennedyOhagan2000,QianWu2008,Gratiet_SIAM_UQ2013}
require hierarchically nested designs and cannot be implemented in
this setting. 

Regarding ABTCK, we consider weakly informative priors with hyper-parameters
$b_{t}=g_{t}=0$, $B_{t}=10$, $\lambda_{t}=2$, $\chi_{t}=2$ and
a mixture prior of Gamma distributions $\phi_{t}|\mathcal{T}\sim0.5\text{G}(1,20)+0.5\text{G}(10,10)$
for $\{\phi_{t}|\mathcal{T}\}$ distributing the prior mass on areas
of smaller and larger values \citep{Gramacy08}. The scale discrepancy
is parametrised as a zero-degree basis expansion $\xi_{k,t}(x|\gamma_{k,t})=\gamma_{k,t}$.
The tree process prior has hyper-parameters $\zeta=0.5$ and $d=2$.
To make the comparison fair, ABCK shares the same settings as ABTCK.
To learn the unknown parameters, we generate a MCMC sample by running
the sampler for $N=25000$ iterations and discarding the first $5000$
sampled values as burn-in.

\begin{figure}[htb!]
\centering \subfloat[\textsf{Prediction GP co-kriging\label{fig:predictR1}}]{\includegraphics[width=0.45\textwidth]{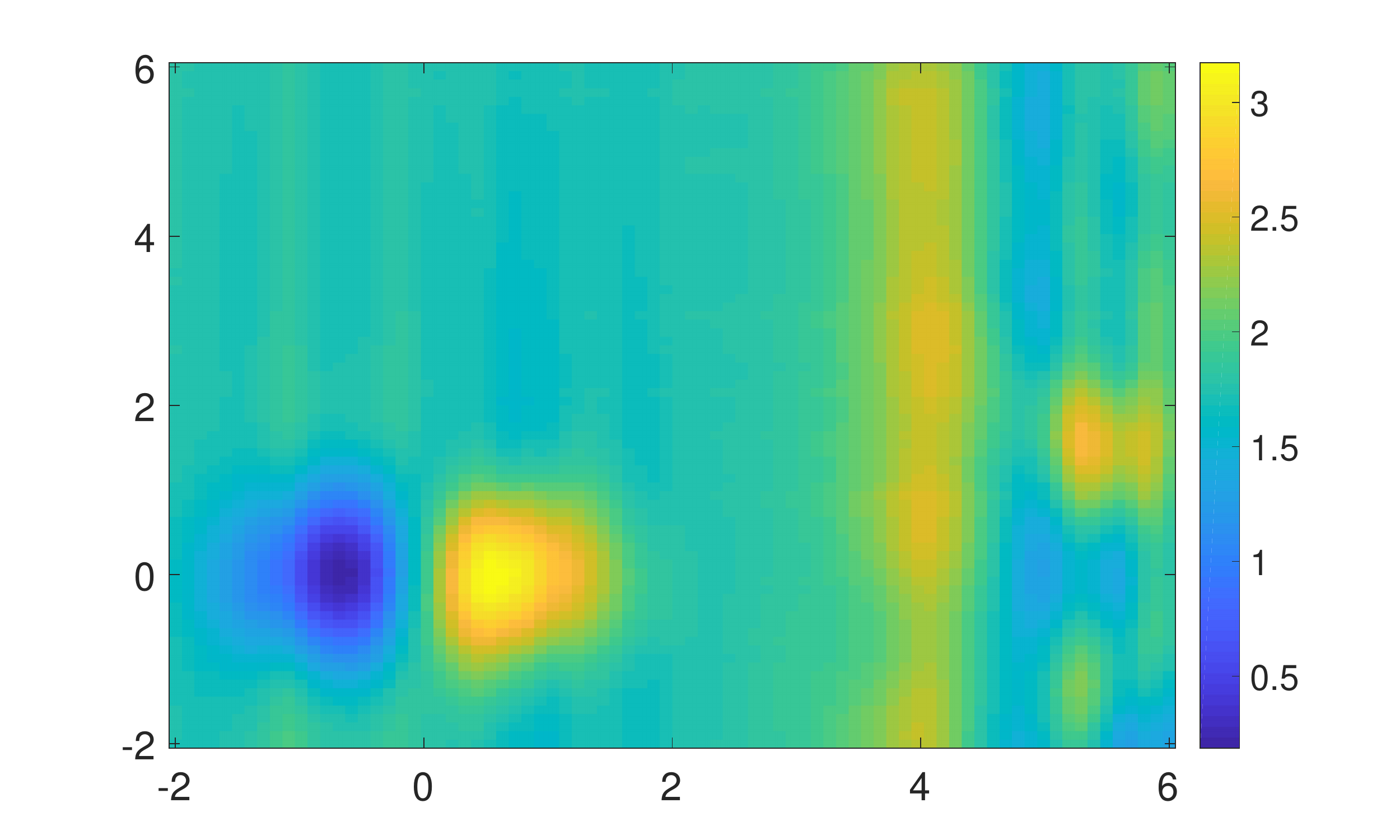}

}\subfloat[\textsf{Prediction Bayesian treed co-kriging\label{fig:predictR1-1}}]{\includegraphics[width=0.45\textwidth]{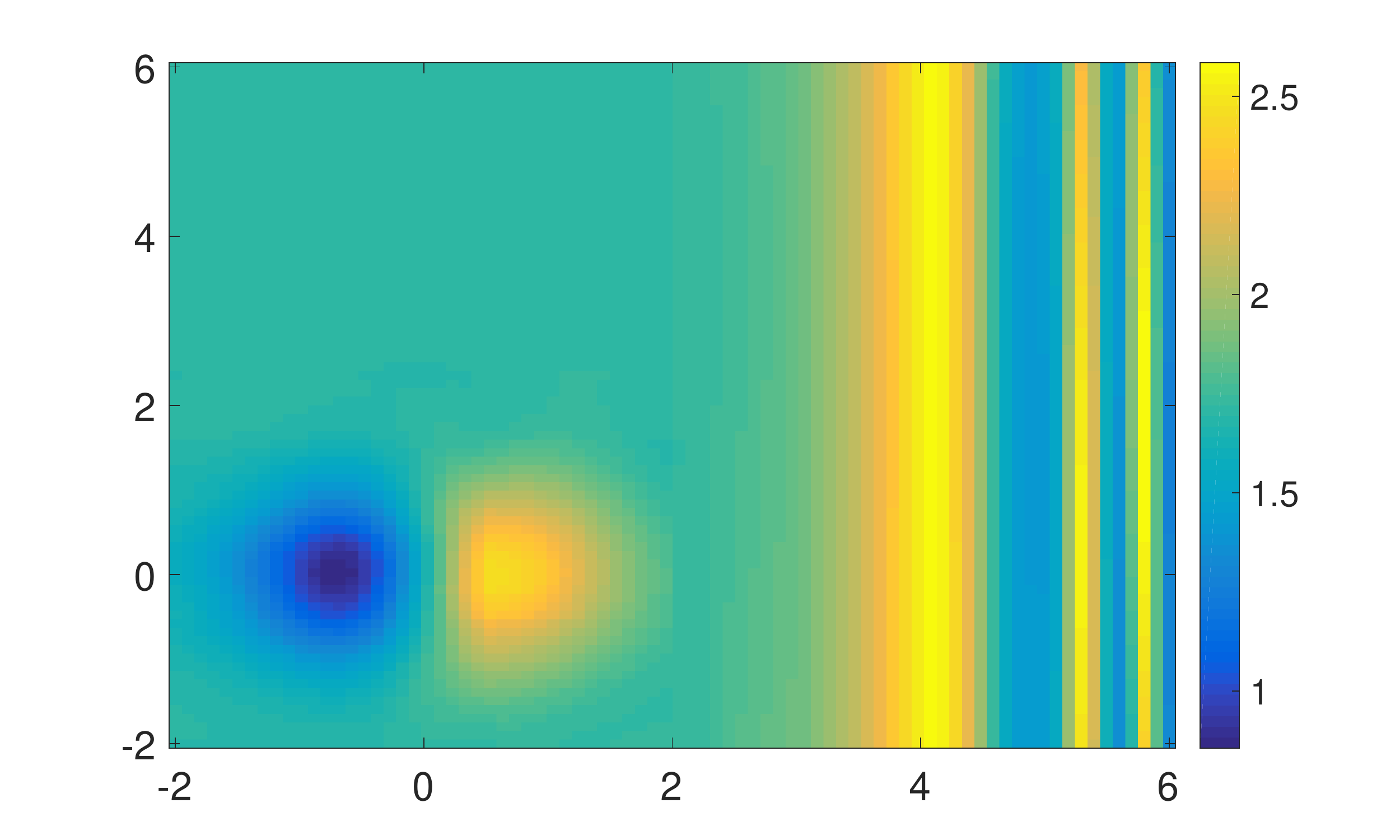}

}

\caption{Prediction of the high-level computer model using two different methods
(a) ABCK and (b) ABTCK.}
\end{figure}

Figures \ref{fig:predictR1} and \ref{fig:predictR1-1} present the
predictive means of $y_{2}(\cdot)$ as functions of the inputs for
ABCK and ABTCK respectively. We observe that the predictive mean produced
by ABTCK is closer to the exact $y_{2}(\cdot)$ than that produced
by ABCK. ABTCK has produced a MSPE $0.0031$ while the stationary
ABCK has produced a MSPE $0.0264$, where MSPE is computed based on
a $100\times100$ grid of input values. This suggests that the treed
partitioning mechanism, as implemented in our ABTCK, is able to successfully
capture and model the non-stationarity, and hence produce more accurate
predictions, in the multifidelity setting. 

The algorithms have been implemented in MATLAB R2017b and run on a
computer with specs: Intel\textsf{Core™i7-7700K CPU @ 4.20GHz $\times$
8, and 62.8 GiB RAM} but in a sequential fashion. The computation
time of training ABTCK was around two times quicker than ABCK. This
is because ABTCK requires the inversion of smaller covariance matrices
than ABCK in the MCMC sampling due to the partitioning. It appears
that the computational overhead introduced by the RJ operation is
dominated by the computational gain due to the partition and subsequent
inversion of smaller matrices.

\begin{figure}[htb!]
\centering \textsf{\includegraphics[width=0.45\linewidth]{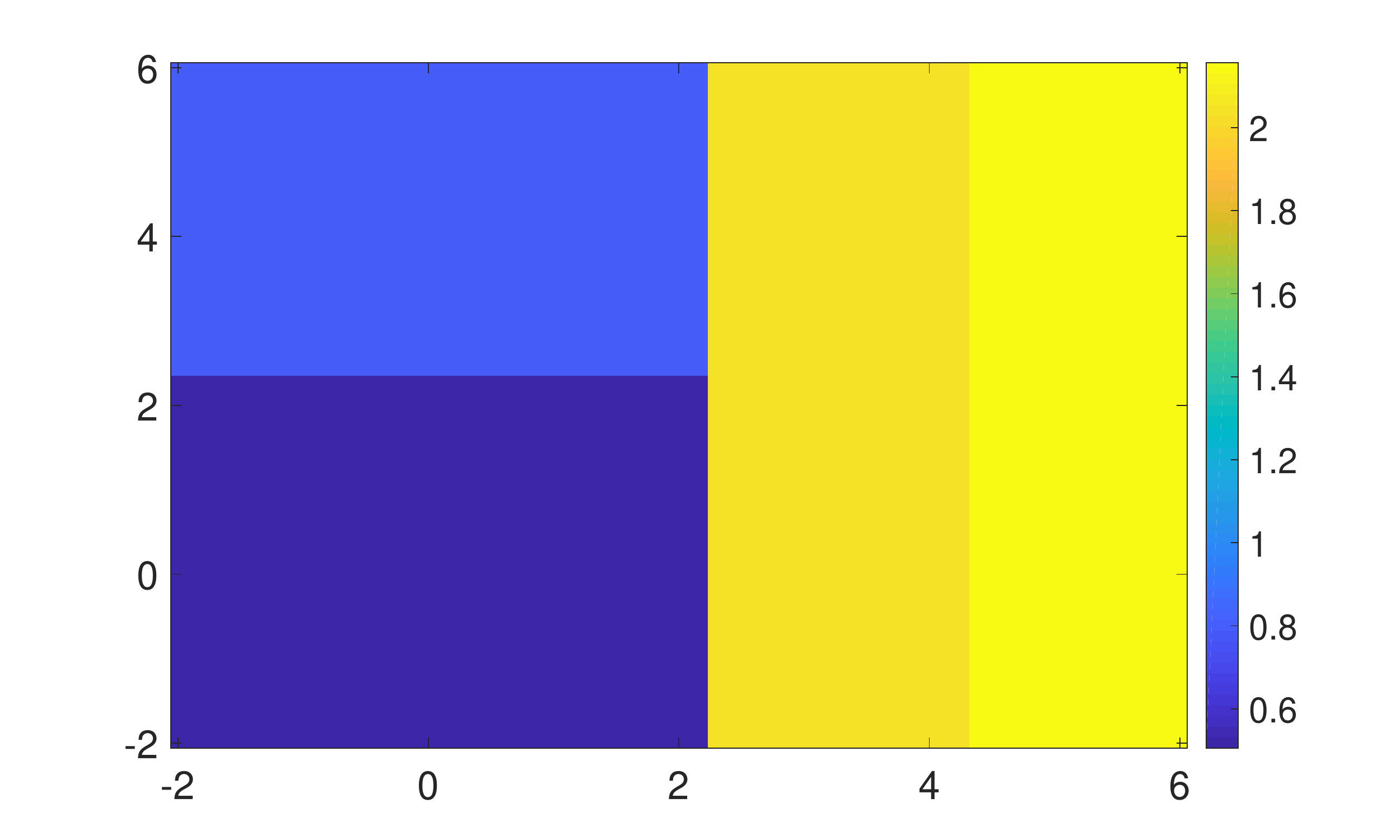}}\caption{Posterior mean of the scalar discrepancy between low and high fidelity
computer models using the augmented Bayesian treed co-kriging.\label{fig:RhoR1}}
\end{figure}

In Figure \ref{fig:RhoR1}, we present the Monte Carlo approximation
of the posterior mean of the scalar discrepancy $\hat{\xi}(x)\approx\frac{1}{N}\sum_{j=1}^{N}w_{t}(x)^{T}\left(\sum_{k=1}^{K^{(i)}}\indicator_{\mathcal{X}_{k}^{(j)}}(x)\hat{\gamma}_{k,t}(\phi_{k,t}^{(j)})\right)$
produced by the ABTCK. We observe that ABTCK has recovered a representation
of the scalar discrepancy which suggests that $\xi(x)$ changes value.
In contrast, ABCK produces a posterior scalar discrepancy which is
equal to $0.525$ and constant throughout the input space due to the
lack of partitioning. 

\section{Application to large-scale climate modeling\label{sec:Application}}

We consider the Advanced Research Weather Research and Forecasting
Version 3.2.1 (WRF Version 3.2.1) climate model \citep{SkamarockKlempDudhiaGillBarkerDudaHuangWangPowers2008}
constrained in the geographical domain $25^{\circ}\text{--}44^{\circ}\text{N}$
and $112^{\circ}\text{--}90^{\circ}\text{W}$ over the Southern Great
Plains (SGP) region, and we concentrate on the average precipitation
response over the area.

We briefly discuss the set-up of the WRF computer model, however more
details can be found in \citep{YanQianLinLeungYangFu2014}. WRF is
employed with the Morrison 2-moment cloud microphysics scheme \citep{morrison2005new}
and the Kain-Fritsch convective parametrisation scheme (KF CPS) \citep{Kain2004}
as in \citep{YangQianLinLeungZhang2012}. The $5$ most critical parameters
\citep{YangQianLinLeungZhang2012,YanQianLinLeungYangFu2014} of the
KF scheme are: the coefficient related to downdraft mass flux rate
$P_{\text{d}}$ that takes values in range $[-1,1]$; the coefficient
related to entrainment mass flux rate $P_{\text{e}}$ that takes values
in range $[-1,1]$; the maximum turbulent kinetic energy in sub-cloud
layer ($m^{2}s^{-2}$) $P_{\text{t}}$ that takes values in range
$[3,12]$; the starting height of downdraft above updraft source layer
(hPa) $P_{\text{h}}$ that takes values in range $[50,350]$; and
the average consumption time of convective available potential energy
$P_{\text{c}}$ that takes values in range $[900,7200]$. The ranges
of the KF CPS parameters are quite wide and hence cause higher uncertainties
in climate simulations due to the non linear interactions and compensating
errors of the parameters \citep{GilmoreStrakaRasmussen2004,MurphyBooth2007methodology,YangQianLinLeungZhang2012}.
We consider the Rapid Radiative Transfer Model (RRTMG) for General
Circulation Models \citep{MlawerTaubmanBrownIaconoClough1997} as
a more accurate radiation scheme for the geological domain of interest.
Here, we are interested in modeling the average precipitation with
respect to the five parameters of the convective parametrisation scheme.

\begin{figure}[htb!]
\centering \subfloat[$\mathscr{C}_{2}:$ $25$ km grid spacing]{\subfloat{\includegraphics[width=0.49\textwidth]{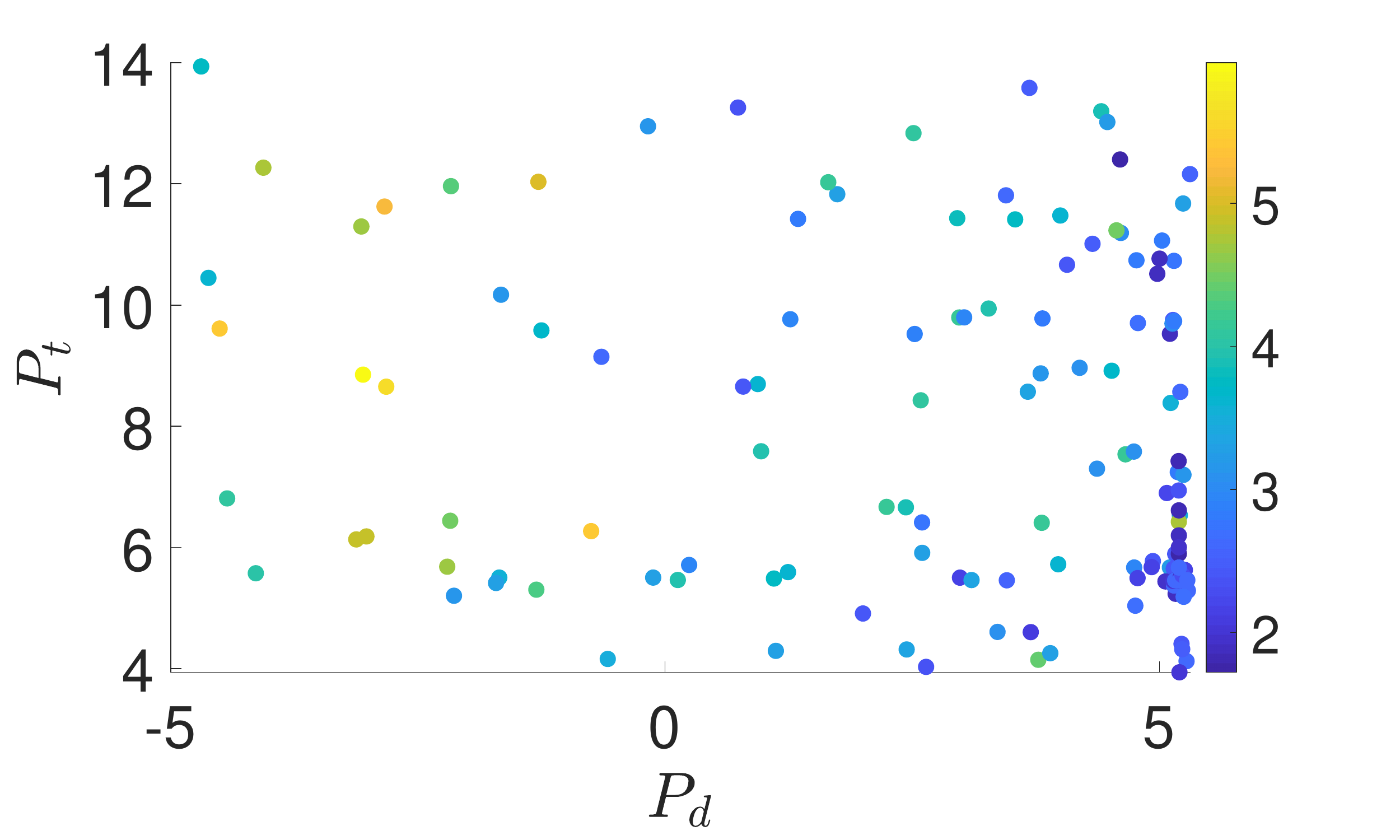}}

}\subfloat[$\mathscr{C}_{1}:$ $12.5$ km grid spacing]{\subfloat{\includegraphics[width=0.49\textwidth]{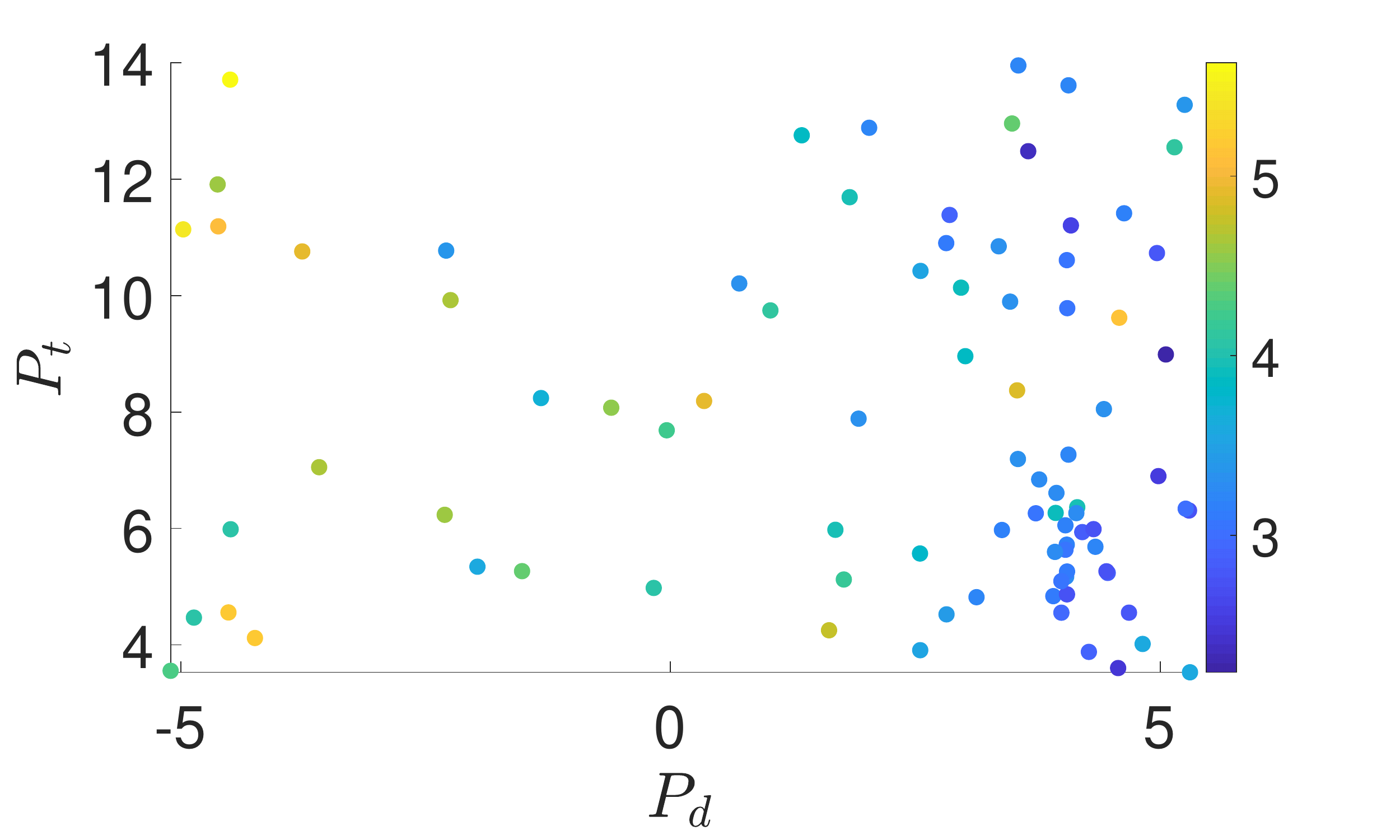}}

}
\caption{Experimental design snapshots \label{fig:data11}}
\end{figure}

The available simulations were generated by running WRF model $240$
times at two resolution levels; $90$ model runs for $12.5$km grid
spacing and $150$ model runs $25$km grid spacing. The fidelity of
the simulations increases when the grid spacing gets finer. The available
simulations have been generated based on a non hierarchically nested
design at the five input parameters (Figure \ref{fig:data11}). The
samples have been generated via a simulated stochastic approximation
annealing (SSAA) calibration algorithm published in \citep{YanQianLinLeungYangFu2014}.
As the SSAA procedure progresses, the sampling range of each parameter
gradually narrows as shown in Figure \ref{fig:data11}. Different
resolutions give different narrowing range on the input space. Due
to the high cost, it is not possible to re-run the expensive WRF model
in order to generate simulations based on a hierarchically nested
design as existing co-kriging methods require. As discussed in \citep{YangQianLinLeungZhang2012,YanQianLinLeungYangFu2014}
the discrepancies between the two fidelity levels may depend on the
five inputs, however no formal statistical analysis have been performed.
The atmospheric humidity at all levels is lower in the fine resolution
than coarse resolution, and the drier atmosphere may result from more
condensation (so more precipitation generated) which consumes more
moisture at the finer resolution. The explicit precipitation increases
with spatial resolution because more clouds are resolved at finer
resolution. Moreover, interest lies in better understanding how different
grid spacing affects the discrepancies in WRF with respect to the
input parameters.

We implement the ABTCK proposed method to analyze the data set. To
make comparisons regarding the necessity of the treed partition as
implemented in our method in the multi-fidelity framework, we consider
the ABCK, namely the ABTCK without the partition mechanism. It is
important to notice that existing co-kriging techniques cannot be
implemented in this application because the available experimental
design is not hierarchically nested. We compare our proposed ABTCK
and ABCK against the standard GP emulator trained against the observed
data of the higher fidelity level only, to demonstrate the importance
of using co-kriging in multi-fidelity problems even under non-hierarchically
nested designs. To ensure fair comparison, the covariance function
family is the same for all three methods, namely: separable square
exponential covariance functions. Regarding the prior model, for the
correlation parameters, we assign Gamma mixture priors $\phi_{k,t}\sim0.5\text{G}(1,10)+0.5\text{G}(5,2)$
distributing the mass on areas of smaller and larger values; for the
binary treed partition priors, we consider hyper-parameters $a=0.8$
and $b=5$; and for the rest parameters we consider weak informative
priors as $b_{t}=0$, $B_{t}=100$, $\lambda_{t}=0.2$, and $\chi_{t}=0.2$.
Regarding the grow \& prune update, we use the prior distributions
as the dimensional matching proposals $\phi_{k,t}\sim0.5\text{G}(1,10)+0.5\text{G}(5,2)$.
We have re-scaled the input space for the five parameters to be between
$[-1,1]$ in order to be able to use the same proposal distribution
for all $\phi_{k,t}$'s. To train the model, we run the MCMC sampler
for $30,000$ iterations from which we discard $5,000$ as burn in.

\begin{figure}[htb!]
\centering {\subfloat{\includegraphics[width=0.65\textwidth]{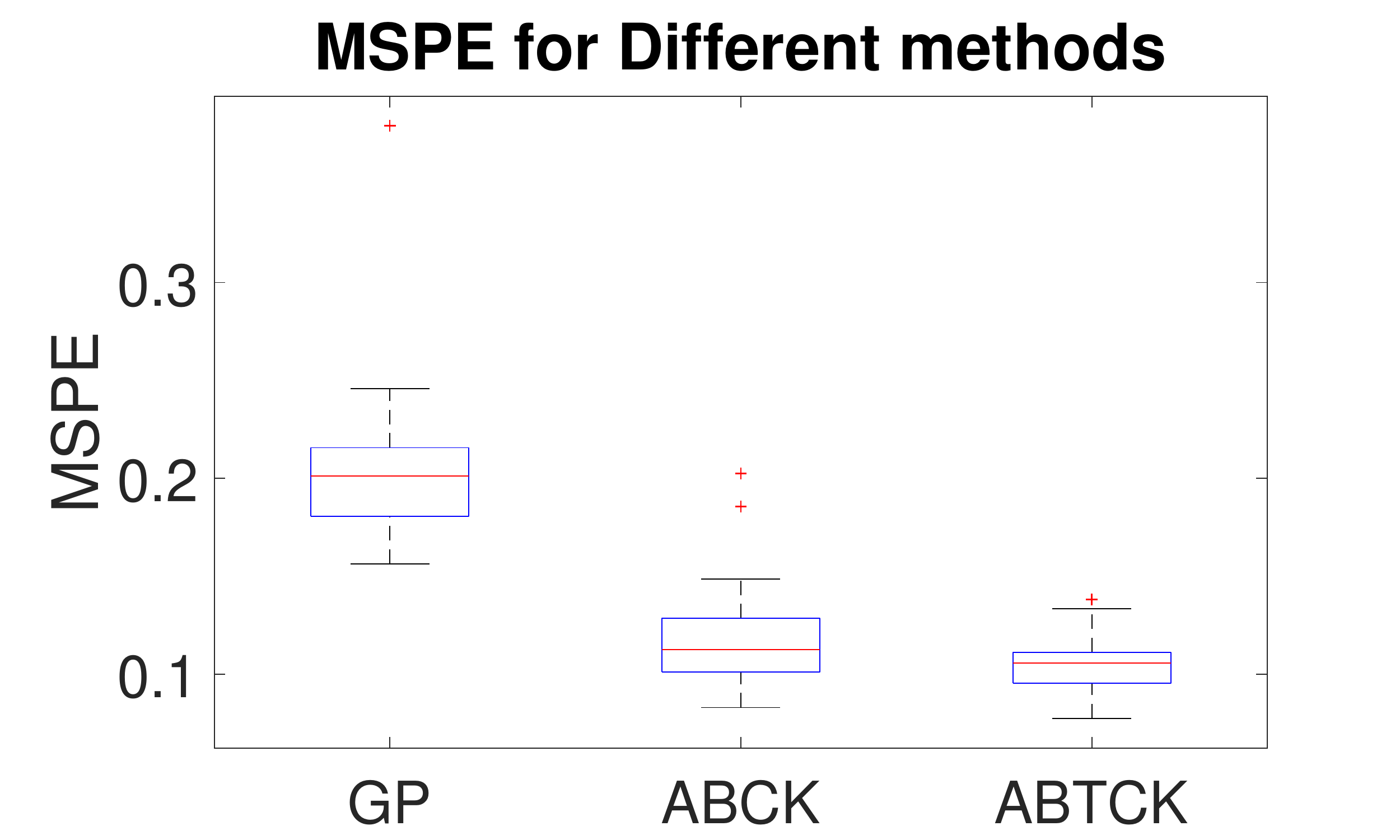}}}
\caption{Boxplot of the MSPE for three different methods \label{fig:MSPE_AP}}
\end{figure}

We randomly choose half of the simulations as the evaluation data-set,
and we use the rest simulations as the training data-set. To account
for the variation due to the stochastic nature of the procedures and
the bias due to the evaluation set, we perform realizations for each
procedure with different evaluation sets each time.The comparison
is performed based on the MSPE, the coverage probability of the $95\%$
equal-tail credible interval (CVG(95\%)), the Nash-Sutcliffe model
efficiency coefficient (NSME), and the computational time. The average
of each of these quantities for the three methods is presented in
Table~1. To give a better representation of the variation, we also
present the boxplots of the MSPEs produced from simple GP, ABCK, and
ABTCK in Figure~\ref{fig:MSPE_AP}.


\begin{table}[htbp]
\centering \caption{\label{table: CV for cokriging} Average of repeated 60 times predictive
performance of three different emulators: Gaussian process, Augmented
Bayesian Co-kriging, Augmented Bayesian Treed Co-kriging}
{\resizebox{1.0\textwidth}{!}{\setlength{\tabcolsep}{2.5em} %
\begin{tabular}{lcccc}
\toprule 
\noalign{\vskip 1.5pt}  & MSPE  & CVG(95\%)  & NSME  & Time(sec) \tabularnewline
\noalign{\vskip 1.5pt} \noalign{\vskip 1.5pt} \noalign{\vskip3pt}
\noalign{\vskip 1.5pt} {GP}  & 0.2118  & 0.613  & 0.31  & 368 \tabularnewline
\noalign{\vskip 4pt} {ABCK}  & 0.1205  & 0.840  & 0.79  & 1804 \tabularnewline
\noalign{\vskip 4pt} {ABTCK}  & 0.0974  & 0.945  & 0.87  & 1240 \tabularnewline
\noalign{\vskip 1.5pt} \bottomrule &  &  &  & \tabularnewline
 &  &  &  & \tabularnewline
\end{tabular}}} 
\end{table}

\begin{figure}[htb!]
\centering \subfloat[\textsf{Real means\label{fig:PFineReal1}}]{\includegraphics[width=0.48\textwidth]{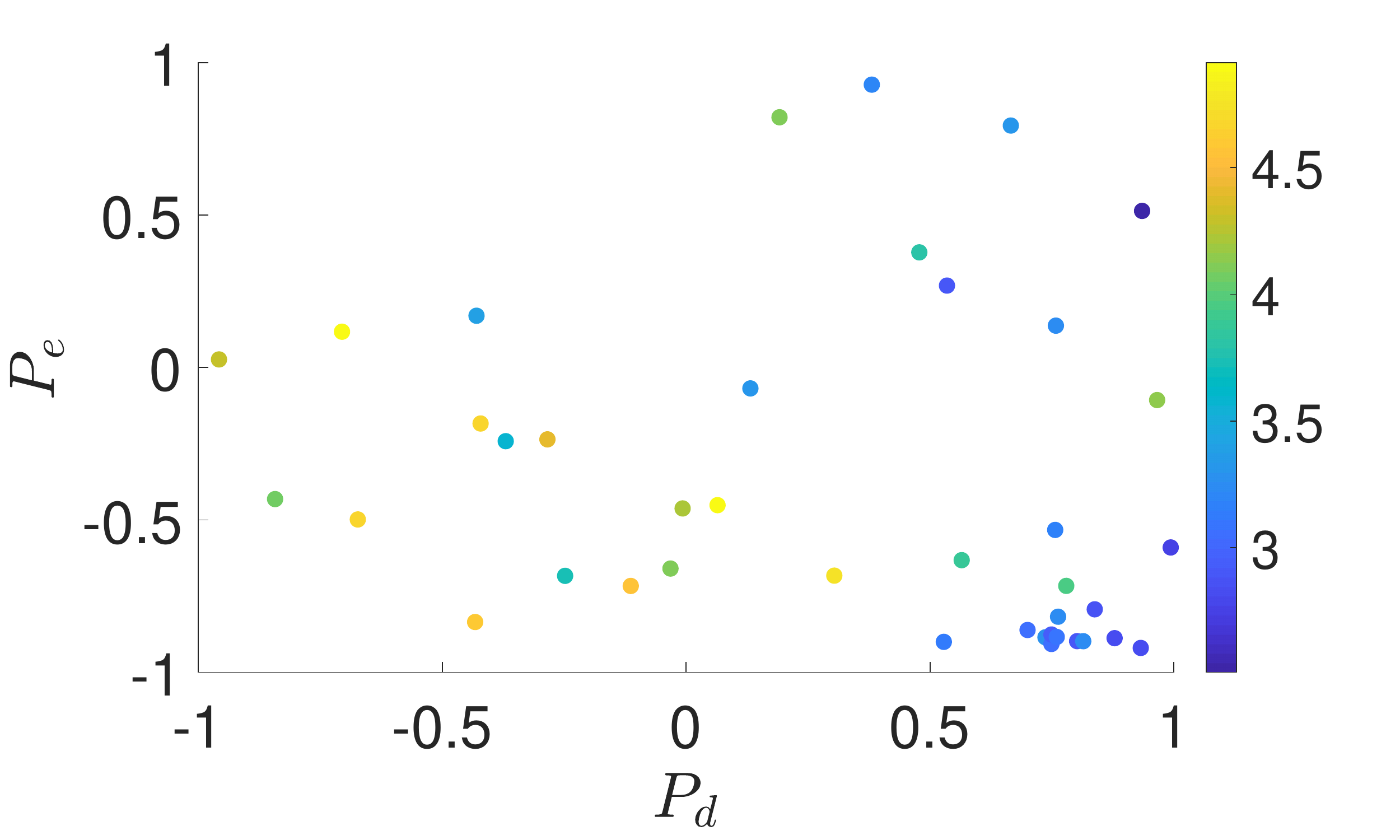}

}\subfloat[GP predictions\textsf{ \label{fig:PFineTGP1}}]{\includegraphics[width=0.48\textwidth]{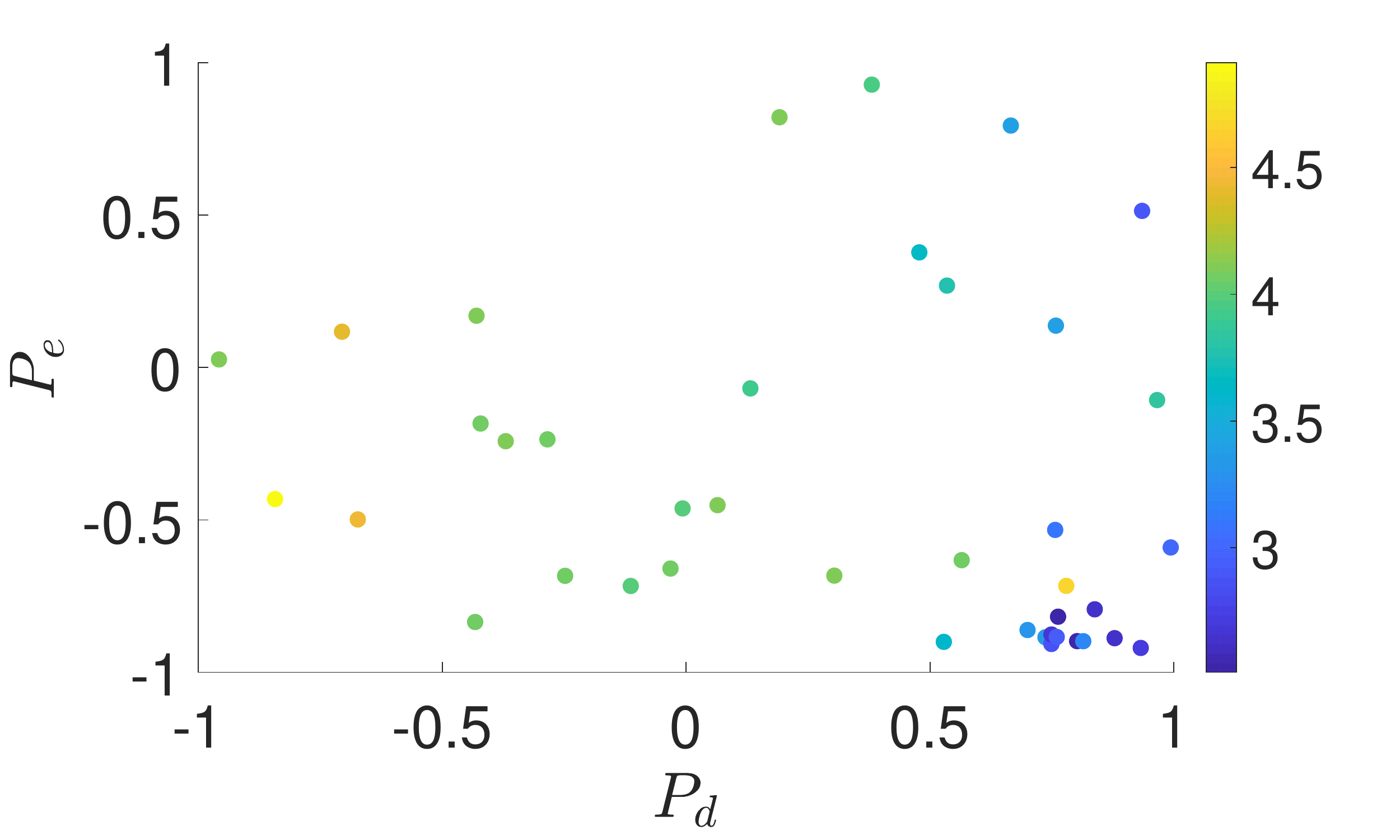}

}\\
\subfloat[ABCK predictions\textsf{\label{fig:PFineGP1}}]{\includegraphics[width=0.48\textwidth]{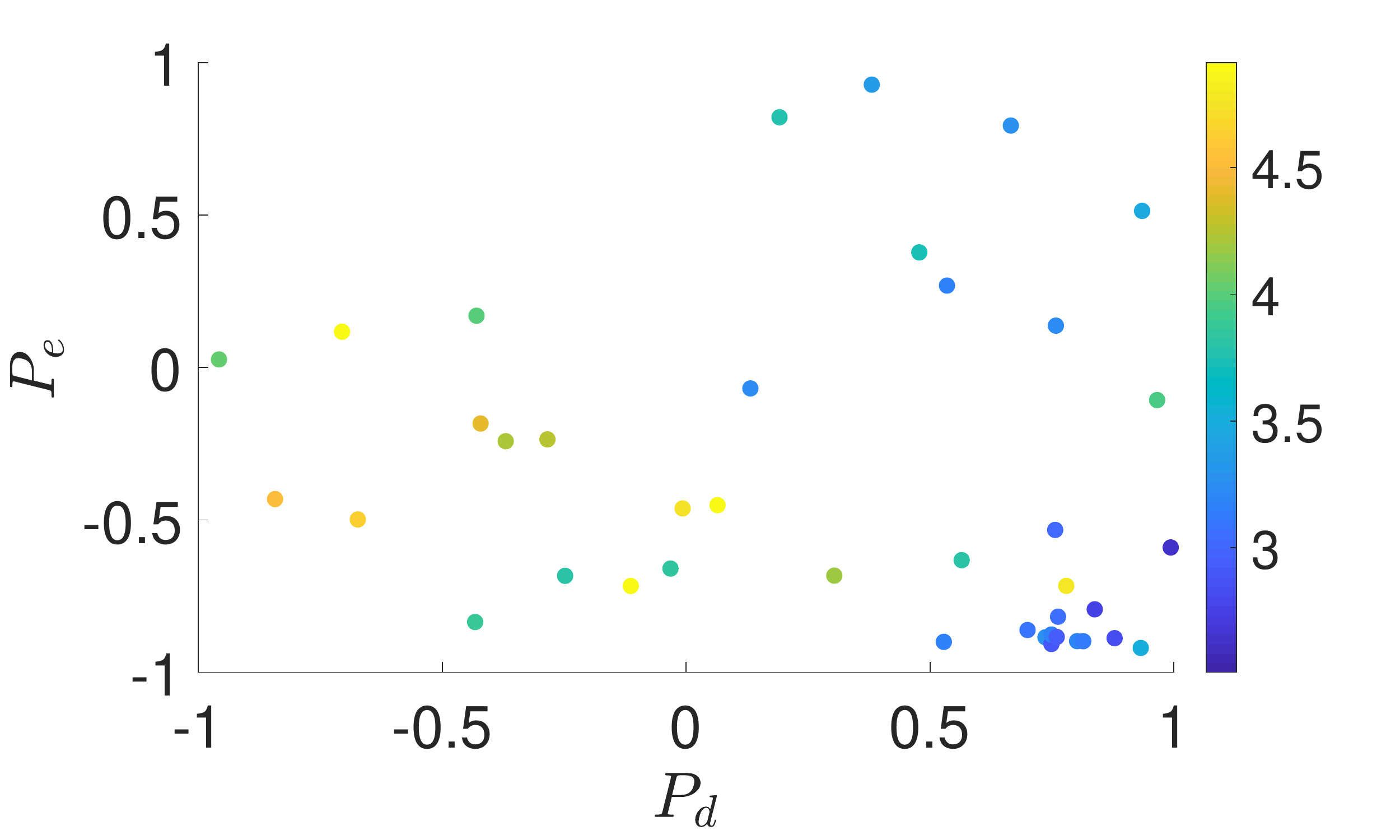}

}
 \subfloat[ABTCK predictions\textsf{ \label{fig:PFineTGP1}}]{\includegraphics[width=0.48\textwidth]{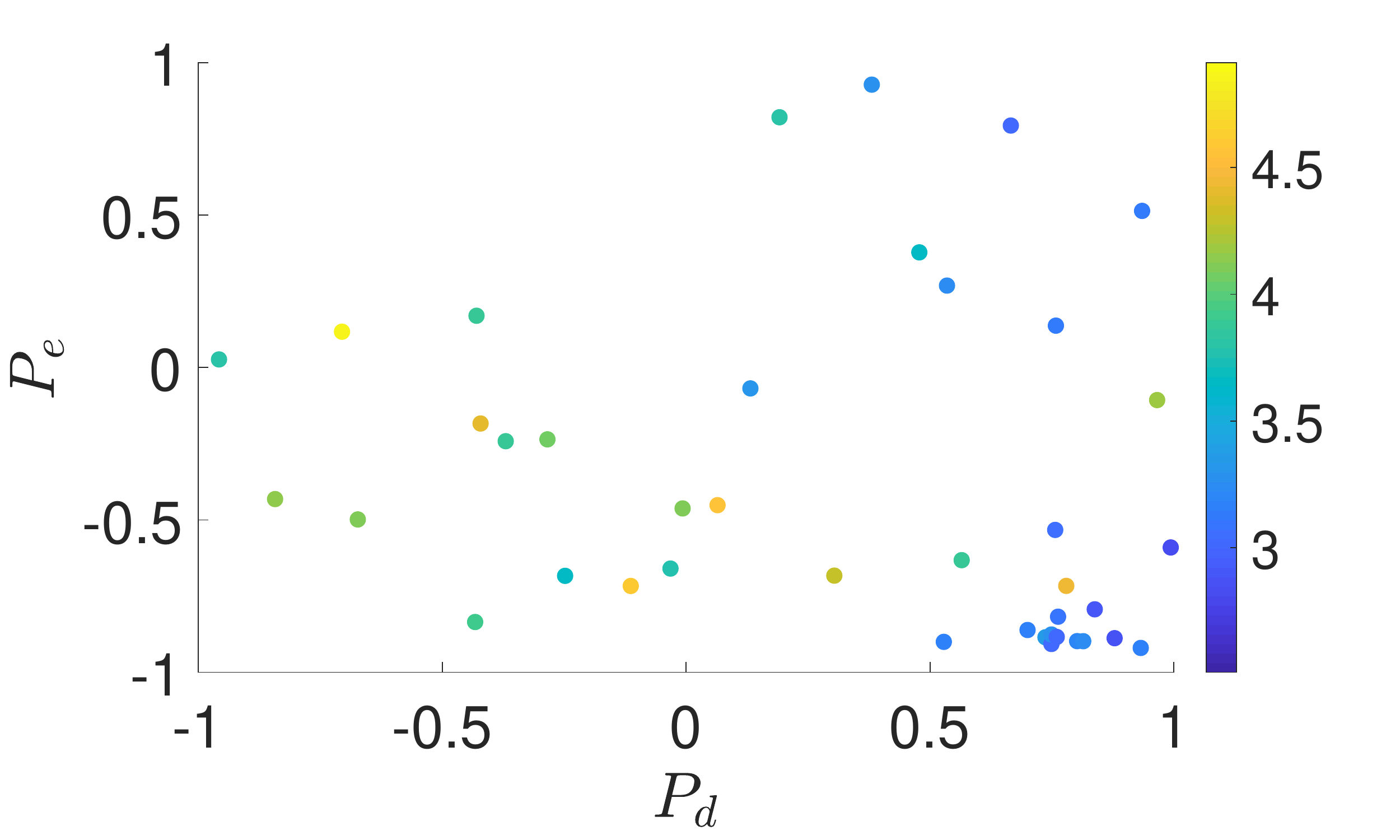}

}

\caption{Real and predicted values produced by ABTCK and ABCK: (a) real realization,
(b) ABCK , (c) ABTCK case 2 .\label{fig:MSPE_AAP}}
\end{figure}

Both ABCK and ABTCK outperform the simple GP by a large margin in
terms of accuracy and constructing more accurate credible intervals.
The mean MSPE and NSME for both ABCK and ABTCK is less than half of
that produced by the simple GP. Moreover, we observe that ABTCK produced
smaller MSPE and NSME than ABCK for all the $60$ realizations, and
hence ABTCK has produced more accurate results than ABCK. The average
MSPE from ABCK is $0.1205$ while the average MSPE from ABTCK is $c$,
which implies an improvement about $20\%$ on the MSPE when we consider
the partition and hence we take into account non-stationarity. The
prediction accuracy is also reflected in the NSME. The average NSME
of the ABTCK is closer to one than both ABCK and GP. Based on the
calculated average CVG(95\%), the ABTCK produced the best representation
of the uncertainty. Not only the ABTCK produced more accurate predictions
but also it gave a better picture of the uncertainty associated with
these predictions. Moreover, the average number of the generated subregions
(tree external nodes) varies from $2$ to $5$. This evidence supports
the use of ABTCK instead of ABCK and hence the use of a non-stationary
process via partitioning. The maximum MSPE difference was $0.0732$
and it was observed in the realization corresponding to the evaluation
dataset (left out simulations) which was more scattered than the rest
in a wider range of the input space. This was almost $60\%$ improvement
in the MSPE. When the majority of the left out simulations are close
to the narrowing range of the simulated input space these differences
become smaller but yet significant. Finally, it is important to notice
that the computational time in ABTCK is approximately two third of
the computational time in ABCK. This means that the improvements on
the prediction and uncertainty described above come in a lower computational
cost. It is worth noticing that we can further reduce the computational
cost of ABTCK if we utilize parallel computing as explained in section~2.5.

In Figure \ref{fig:MSPE_AAP}, we plot the simulated precipitation
from WRF at high fidelity, the predicted average precipitation produced
from ABTCK, from ABCK, and from simple GP with respect to the downdraft
mass flux rate $P_{\text{d}}$ and the coefficient related to entrainment
mass flux rate $P_{\text{e}}$. Precisely, we present the case corresponding
to realizations with the highest MSPE differences between ABTCK and
GP. It is obvious that the GP is not able to capture the variation
in the central part of the plot where observations for high level
model are sparse. Both ABCK and ABTCK are able to capture that variation
with the help from the low fidelity model. regarding the differences:
we observe that ABCK produced a smoother representation of the precipitation,
however ABTCK was able to more accurately represent the local features.
This is especially noticeable on the middle of Figure \ref{fig:MSPE_AAP}.
The prediction is much improved over the whole left out simulation
runs even in the clustered small range.


\section{Conclusions and further work \label{sec:Conclusions-and-further}}

We built a Bayesian emulator for the Weather Research and Forecasting
(WRF) model. The proposed method, called Augmented Bayesian Treed
Autoregressive Co-Kriging, extends the scope of the co-kriging methods.
First our procedure can be implemented in problems where the experimental
design is not necessarily hierarchically nested while keeping the
computational demands low. This overcomes the difficulty of existing
co-kriging methods which require hierarchically nested designs in
order to keep the computations practically feasible. Secondly, our
method can account for non-stationarity, and potential discontinuity,
in the output of the computer models without the need to specify complicated
or problem specific GP priors, in the multifidelity setting. Finally,
we propose the use of a Monte Carlo recursive emulator which can recover
the predictive distribution of the computer model output at every
level, and can be used with non-hierarchically nested designs as well,
while keeping the computational cost lower than the existing emulators
as it requires operations with smaller matrices.

We analyzed the Weather Research and Forecasting (WRF) simulator using
the Kain-Fritsch convective parametrisation scheme by using our novel
procedure. This is a large-scale climate modeling application where
the available simulations are performed at different fidelity levels
at non hierarchically nested designs. Our method discovered non-stationarity
in the WRF output precipitation with respect to the KFC input parameters.
We observed that the use of Bayesian treed partition in the co-kriging
framework as utilized in our method is able to provide more accurate
predictions than ignoring it. For instance, in the WRF application
we observed the use of the partition was able to reduce the MSPE around
$21\%$ on average when we compared the ABTCK with the ABCK where
the partitioning was dropped out. In our simulation example considering
non-nested designs, we observed that the augmentation mechanism was
able to recover the model output accurately enough. 

The procedure can be modified to involve a basis selection mechanism
for $h_{t}(\cdot)$ of $\{\delta_{t}(x)\}$ and $w_{t}(\cdot)$ of
$\{\xi_{t}(\cdot)\}$ at different input sub-regions $\mathcal{X}_{k,t}$,
by properly specifying spike-and-slab priors on $\beta_{k,t}$ and
$\gamma_{k,t}$ and calculating Gibbs updates. One can use the fixed
hyper-parameters of the latent treed process $\pi(\mathcal{T})$ to
control or mitigate possible non-identifiability between the discrepancy
functions, by setting $\xi_{k,t}(x)=\gamma_{k,t}$ and meaningful
priors on $\delta_{k,t}(\cdot)$ in the sense of \citep{brynjarsdottir2014learning}.
The rational is that the treed prior can act as a penalty favoring
simpler partitions, which can mitigate the competition between the
two discrepancies. An extension of ABTCK would be to specify different
partitions for $\xi_{t}(x)$, $\delta_{t}(x)$, $y_{1}(x)$, which
may lead to a more flexible model, however, it is not clear if conditional
posteriors can still be marginalized to keep the computational demands
feasible. The authors are currently working on a sequential design
procedure with multifidelity simulations that take into account non-hierarhically
nested designs.

\begin{singlespace}
 \bibliographystyle{asa}
\bibliography{main}

\end{singlespace}

\appendix

\section{Appendix\label{sec:Appendix}}

Let $\mathfrak{Z}$, $\mathfrak{J}$ denote any sub-sets of the design
$\mathfrak{\tilde{X}}_{t}$ for $t=1,...,S$. Let $|\mathfrak{Z}|$
denote the size of $\mathfrak{Z}$, and let $y_{0}(\cdot)=0$ and
$\xi_{0}(\cdot)=0$. The parameters of the conditional distributions
in \eqref{eq:pouitsesmple}-\eqref{eq:antegeiametakitsoukala} are
\begin{align}
\hat{B}_{t}(\phi|\mathfrak{Z})= & [H_{t}^{\top}(\mathfrak{Z})R_{t}^{-1}(\mathfrak{Z},\mathfrak{Z}|\phi)H_{t}(\mathfrak{Z})+B_{t}^{-1}]^{-1},\;t=1:S\label{eq:trfjhfghdfh}\\
\hat{\beta}_{t}(\phi|\mathfrak{Z})= & \hat{B}_{t}(\phi|\mathfrak{Z})[H_{t}^{\top}(\mathfrak{Z})R_{t}^{-1}(\mathfrak{Z},\mathfrak{Z}|\phi)[y_{t}(\mathfrak{Z})-\xi_{t-1}(\mathfrak{Z}|\gamma_{t-1})\circ y_{t-1}(\mathfrak{Z})]+B_{t}^{-1}b_{t}]\label{eq:sfgbxcvbzcbh}\\
\hat{G}_{t-1}(\phi|\mathfrak{Z})= & [W_{t-1}(\mathfrak{Z};y_{t-1})C_{t-1}(\phi|\mathfrak{Z})W_{t-1}^{\top}(\mathfrak{Z};y_{t-1})+G_{t-1}^{-1}]^{-1},\;t=2:S\nonumber \\
\hat{\gamma}_{t-1}(\phi|\mathfrak{Z})= & \hat{G}_{t-1}(\phi|\mathfrak{Z})[G_{t-1}^{-1}g_{t-1}+W_{t-1}^{\top}(\mathfrak{Z};y_{t-1})\hat{C}_{t-1}(\phi|\mathfrak{Z})[y_{t}(\mathfrak{Z})-H_{t}(\mathfrak{Z})b_{t}]],\,t=2,...,S\nonumber 
\end{align}
\begin{align}
\hat{C}_{t-1}(\phi|\mathfrak{Z})= & R_{t}^{-1}(\mathfrak{Z},\mathfrak{Z}|\phi)+R_{t}^{-1}(\mathfrak{Z},\mathfrak{Z}|\phi)H_{t}(\mathfrak{Z})\nonumber \\
 & \qquad\times[R_{t}^{-1}(\mathfrak{Z},\mathfrak{Z}|\phi)+H_{t}(\mathfrak{Z})B_{t}^{-1}H_{t}^{\top}(\mathfrak{Z})]H_{t}^{T}(\mathfrak{Z})R_{t}^{-1}(\mathfrak{Z},\mathfrak{Z}|\phi),\;t=2,...,S\nonumber \\
\hat{\lambda}_{t}(\mathfrak{Z})= & \lambda_{t}+\frac{|\mathfrak{Z}|}{2},\;t=1,...,S\nonumber \\
\hat{\chi}_{t}(\phi|\mathfrak{Z})= & (|\mathfrak{Z}|+2\lambda_{t}-2)\hat{\sigma}_{t}^{2}(\phi|\mathfrak{Z}),\;t=1,...,S\nonumber \\
\hat{\sigma}_{t}^{2}(\phi|\mathfrak{Z})= & \frac{1}{2\lambda_{t}+|\mathfrak{\mathfrak{Z}}|-2}\left(2\chi_{t}+y_{t}^{\top}(\mathfrak{\mathfrak{Z}})R_{t}^{-1}(\mathfrak{Z},\mathfrak{Z}|\phi)y_{t}(\mathfrak{Z})+b_{t}^{\top}B_{t}^{-1}b_{t}\right.\nonumber \\
 & \qquad\left.+g_{t-1}^{\top}G_{t-1}^{-1}g_{t-1}-\hat{\alpha}_{t}^{\top}(\phi,\mathfrak{Z})\hat{A}_{t}^{-1}(\phi|\mathfrak{Z})\hat{\alpha}_{t}(\phi|\mathfrak{Z})\right),\;t=1,...,S\label{eq:dghdfghdafsg}\\
\hat{A}_{t}(\phi|\mathfrak{Z})= & \left[L_{t}(\mathfrak{Z};y_{t-1})^{\top}R_{t}^{-1}(\mathfrak{Z},\mathfrak{Z}|\phi)L_{t}(\mathfrak{Z};y_{t-1})+\text{diag}(B_{t}^{-1},G_{t-1}^{-1})\right]^{-1}\,;\label{eq:wthtg}\\
\hat{\alpha}_{t}(\phi|\mathfrak{Z})= & \hat{A}_{t}(\phi|\mathfrak{Z})\left(L_{t}(\mathfrak{Z};y_{t-1})^{\top}R_{t}^{-1}(\mathfrak{Z},\mathfrak{Z}|\phi)+\left[b_{t}^{\top}B_{t}^{-1},g_{t-1}^{\top}G_{t-1}^{-1}\right]^{\top}\right).\label{eq:htwhyetgy}
\end{align}
where: $W_{t-1}(\mathfrak{Z};y_{t-1})=\text{diag}(y_{t-1}(\mathfrak{Z}))w_{t-1}(\mathfrak{Z})$
for $t=2,...,S$ and $W_{0}(\mathfrak{Z};\cdot)=0$; $L_{t}(\mathfrak{Z};y_{t-1})=\left[H_{t}(\mathfrak{Z}),\text{diag}(y_{t-1}(\mathfrak{Z})W_{t-1}(\mathfrak{Z}))\right]$
for $t=2,...,S$ and $L_{1}(\mathfrak{Z};\cdot)=H_{1}(\mathfrak{Z})$.
In the manuscript, when $\mathfrak{Z}=\mathfrak{X}_{k,t}$, we use
notation $\hat{B}_{k,t}=\hat{B}_{t}(\phi|\tilde{\mathfrak{X}}_{k,t})$,
$\hat{\beta}_{t}(\phi)=\hat{\beta}_{t}(\phi|\tilde{\mathfrak{X}}_{k,t})$,
etc... to facilitate the presentation.

The equations of the functions $\hat{R}_{k,t}$, $\hat{\mu}_{(t-1)\rightarrow t}$,
and $\hat{\mu}_{(t+1)\rightarrow t}$ in \eqref{eq:sdjksdbfg} 
\begin{align}
\hat{R}_{t}(\phi|\mathfrak{Z};\mathfrak{J})= & R_{t}(\mathfrak{Z},\mathfrak{Z}|\phi)-R_{t}(\mathfrak{Z},\mathfrak{J}|\phi)R_{t}^{-1}(\mathfrak{J},\mathfrak{J}|\phi)R_{t}^{\top}(\mathfrak{Z},\mathfrak{J}|\phi)\nonumber \\
 & \qquad+\left[H_{t}(\mathfrak{Z})+R_{t}(\mathfrak{Z},\mathfrak{J}|\phi)R_{t}^{-1}(\mathfrak{J},\mathfrak{J}|\phi)H_{t}(\mathfrak{J})\right]\hat{B}_{t}(\phi|\mathfrak{J})\nonumber \\
 & \qquad\qquad\qquad\qquad\times\left[H_{t}(\mathfrak{Z})+R_{t}(\mathfrak{Z},\mathfrak{J}|\phi)R_{t}^{-1}(\mathfrak{J},\mathfrak{J}|\phi)H_{t}(\mathfrak{J})\right]^{\top}\label{eq:etywsywet}\\
\hat{\mu}_{(t-1)\rightarrow t}(\phi,\gamma|\mathfrak{Z};\mathfrak{J})= & \xi_{t-1}(\mathfrak{Z}|\gamma)\circ y_{t-1}(\mathfrak{Z})+H_{t}(\mathfrak{Z})\hat{\beta}_{t}(\phi|\mathfrak{J})\nonumber \\
 & \qquad\qquad+R_{t}(\mathfrak{Z},\mathfrak{J}|\phi)R_{t}^{-1}(\mathfrak{J},\mathfrak{J}|\phi)\nonumber \\
 & \qquad\qquad\qquad\times\left[y_{t}(\mathfrak{J})-\xi_{t-1}(\mathfrak{J}|\gamma)\circ y_{t-1}(\mathfrak{J})-H_{t}(\mathfrak{J})\hat{\beta}_{t}(\phi|\mathfrak{J})\right],\,t=1:S\nonumber \\
\hat{\mu}_{(t+1)\rightarrow t}(\phi,\gamma|\mathfrak{Z};\mathfrak{J})= & y_{t+1}(\mathfrak{Z})-H_{t+1}(\mathfrak{Z})\hat{\beta}_{t+1}(\phi|\mathfrak{J})\nonumber \\
 & \qquad\qquad-R_{t+1}(\mathfrak{Z},\mathfrak{J}|\phi)R_{t+1}^{-1}(\mathfrak{J},\mathfrak{J}|\phi)\label{eq:wtytywet}\\
 & \qquad\qquad\qquad\times\left[y_{t+1}(\mathfrak{J})-\xi_{t}(\mathfrak{J}|\gamma)\circ y_{t}(\mathfrak{J})-H_{t+1}(\mathfrak{J})\hat{\beta}_{t+1}(\phi|\mathfrak{J})\right],\,t=1:S-1\nonumber 
\end{align}

\part*{Supplementary material }

\section{Heat transfer example}

We examine the modeling and predictive benefits of introducing the
binary treed partition mechanism in the Bayesian co-kriging setting,
when the experimental design is hierarchically nested. So we compare
the proposed\textit{ Bayesian treed co-kriging (BTCK)} method (imputation
mechanism is doped out here) against the existing co-kriging model.
The procedures were implemented in MATLAB R2017b, and ran on a computer
with specifications (Intel\textsf{Core™i7-7700K CPU @ 4.20GHz $\times$
8, and 62.8 GiB RAM)}. 

We consider the benchmark problem of a heated metal block with a rectangular
cavity, which can be modeled as an elliptic partial differential equation.
Assume that there are three computer models aiming at describing the
steady state of the temperature, and they are arranged in ascending
order of fidelity as $\{\mathscr{C}^{(t)}\}_{t=1}^{3}$.

\begin{figure}[htb!]
\centering

\subfloat[\textsf{Fine grid \label{fig:Fine}}]{\includegraphics[width=0.45\textwidth]{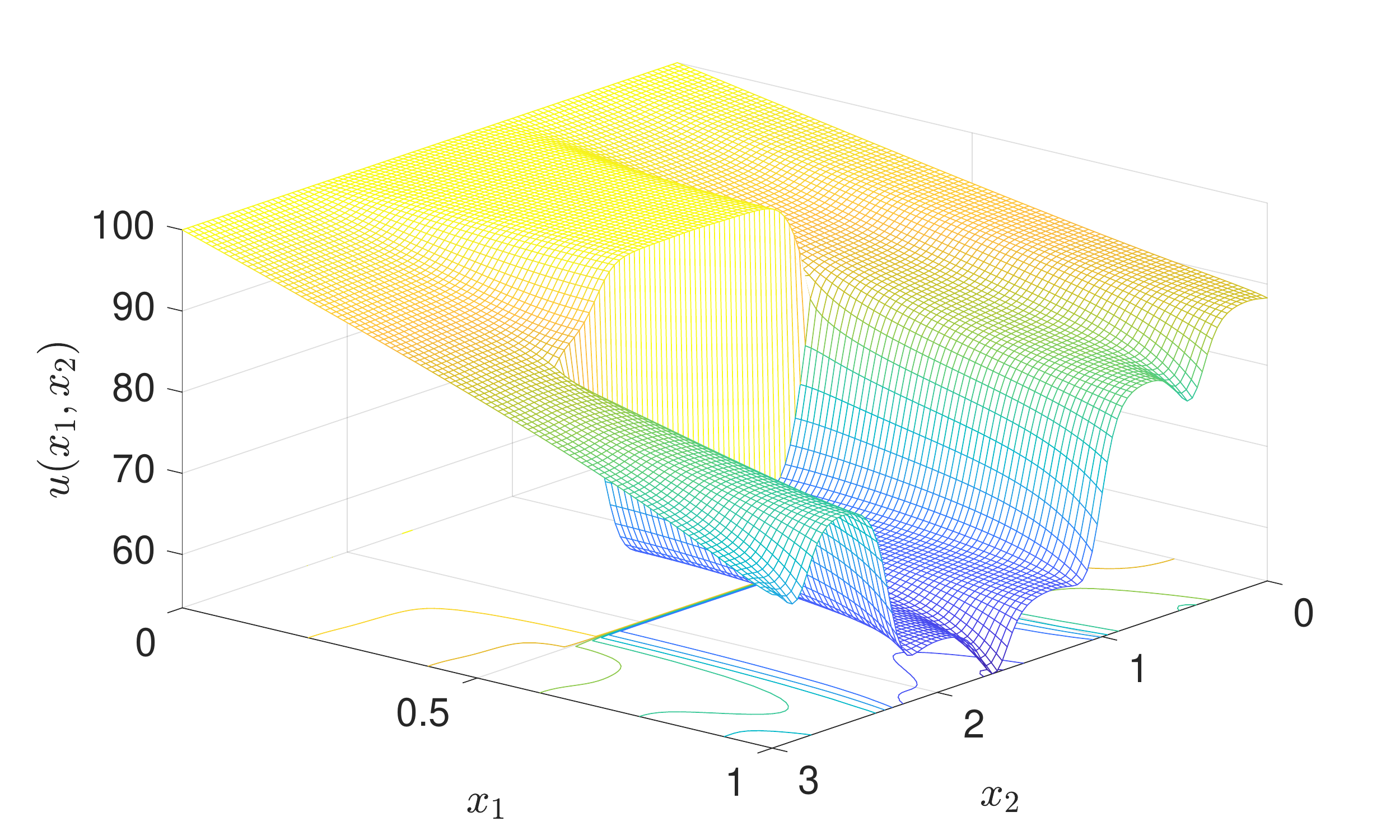}

}\subfloat[\textsf{Intermidiate grid\label{fig:Inter}}]{\includegraphics[width=0.45\textwidth]{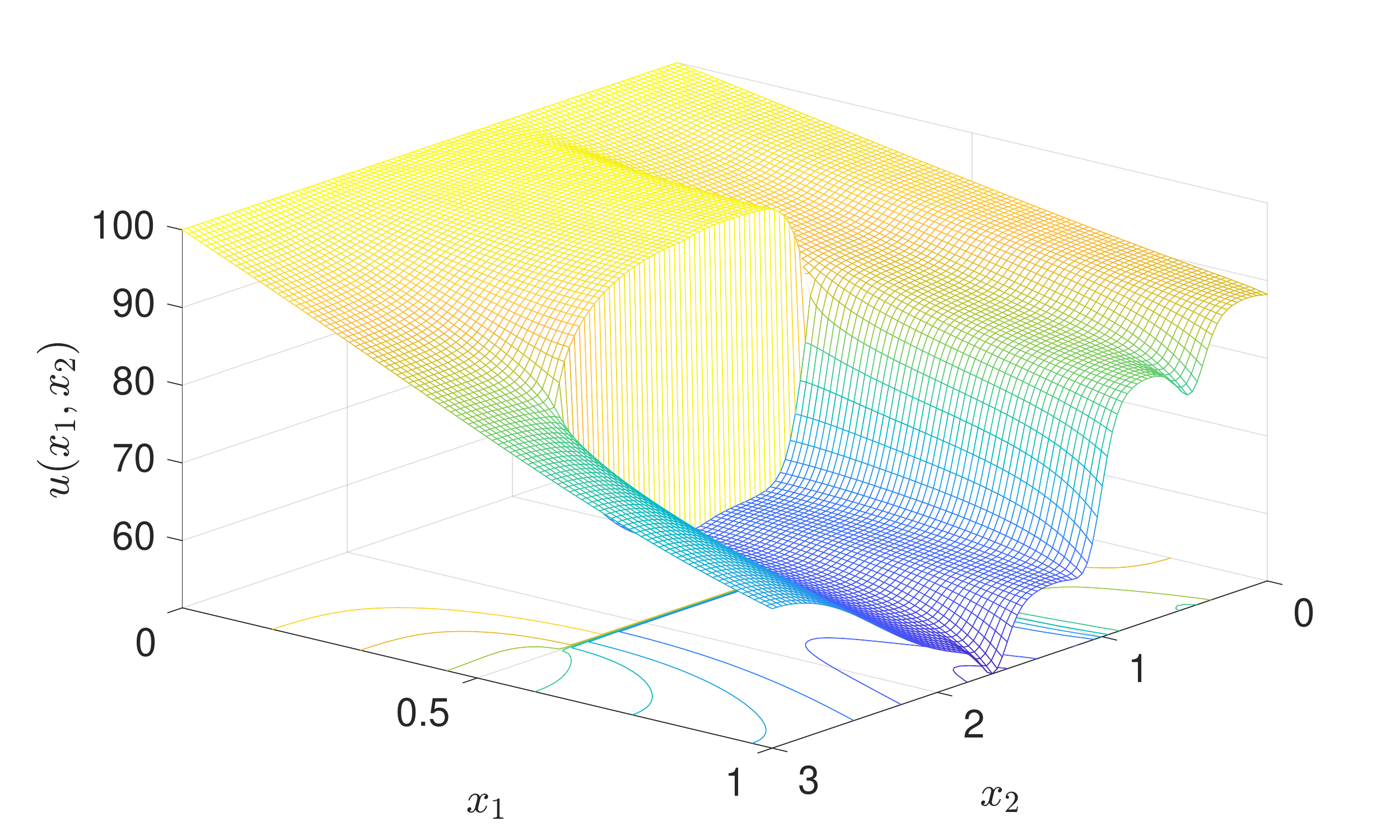}

}

\subfloat[\textsf{Corse grid \label{fig:Corse}}]{\includegraphics[width=0.45\textwidth]{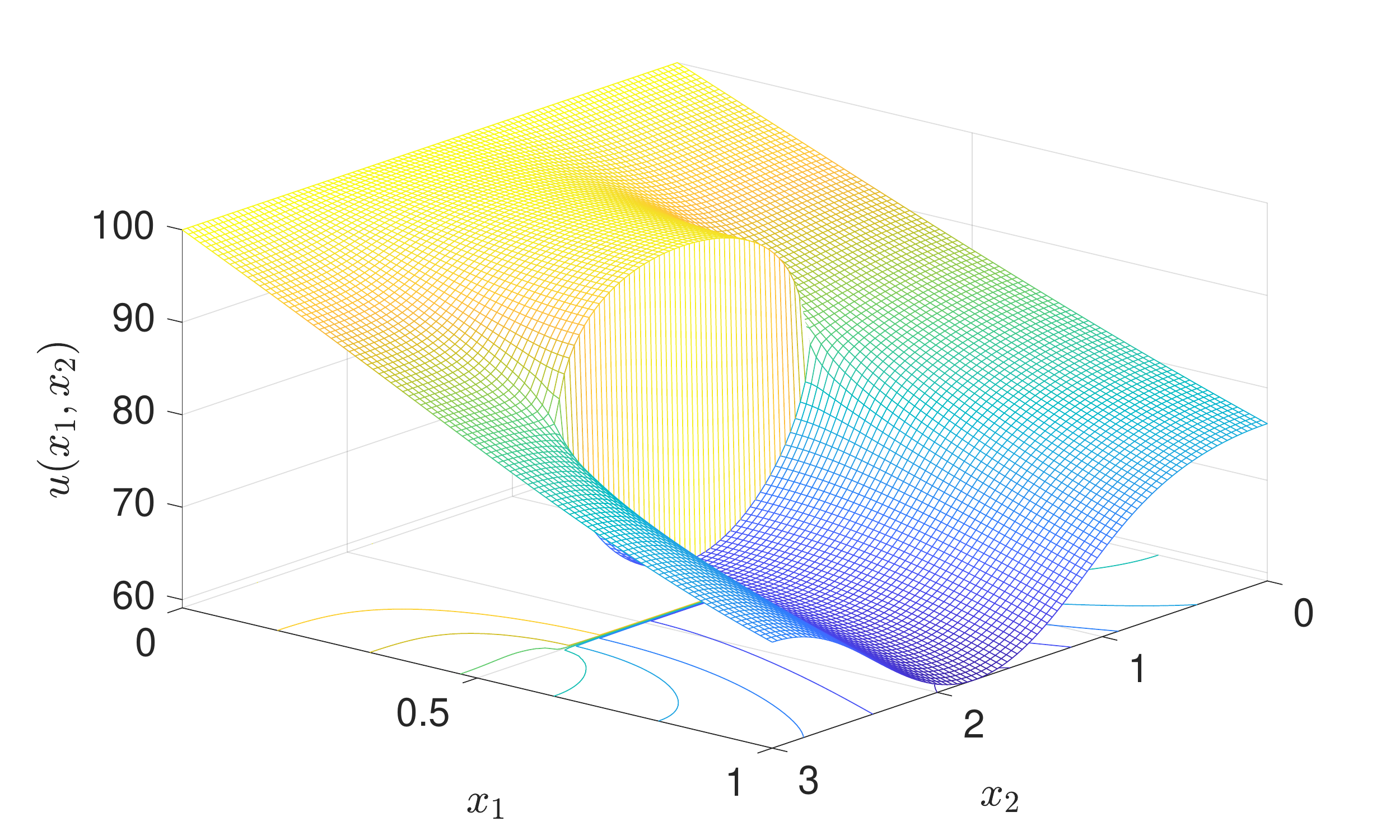}

}\subfloat[\textsf{Fine grid \label{fig:Fine-1}}]{\includegraphics[width=0.45\columnwidth]{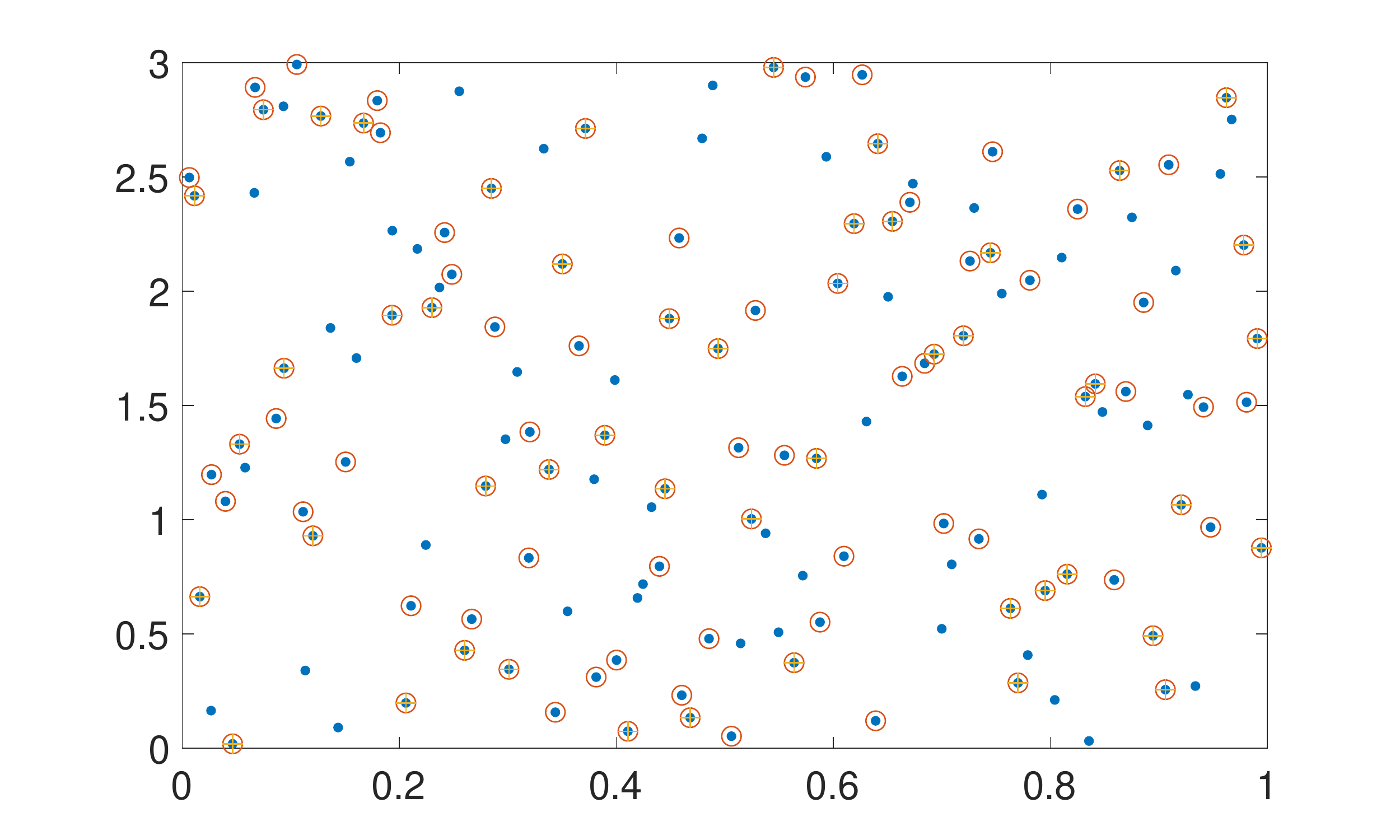}

}

\caption{Response surface for the temperature, steady state solution at three
levels of accuracy and sampling design: (a) Coarse computer model,
(b) Intermediate computer model, and (c) Fine computer model, and
(d) sampling design.\label{fig:theta_histR1}}
\end{figure}

Let us consider 2D elliptic PDEs 
\begin{align}
-\nabla\cdot c^{(j)}(x)\nabla u^{(j)}(x) & =f(x),\hfill\label{eq:spde_ex1_1}
\end{align}
for $x\in\mathcal{X}-\partial\mathcal{X}$ where $x=(x_{1},x_{2})$,
that describes a rectangular block of size $\mathcal{X}=[0,1]\times[0,3]$,
with a rectangular cavity of size $[0.5,0.015]\times[1,2.5]$. We
consider that the left side of the block is heated to $100$ degrees
and hence we consider Dirichlet condition $u=100$. At the right side
of the metal block, heat is flowing from the block to the surrounding
air at a constant rate and we assume Neumann condition $\frac{\partial u}{\partial n}=-20$.
The rest boundary conditions are Neumann condition $\frac{\text{d}}{\text{d}n}u=0$.
The internal heat source is $f(x)=1$. The spatial dependent thermal
connectivity is denoted as $c^{(j)}(x)$; it is $c^{(1)}(x)=1$ for
the least accurate computer model, $c^{(2)}(x)=\exp(1.5\sin(3.33\pi x_{2}))\indicator(x_{2}<1.8)$
for more accurate computer model, and $c^{(3)}(x)=\exp(1.5\sin(3.33\pi x_{2}))$
for most accurate computer model. The PDE in \eqref{eq:spde_ex1_1}
is solved via a FEM solver with the domain $\mathcal{X}$ discretized
in $24119$ nodes. 
We are interested in recovering the temperature $u(x)$, in the steady
state. The temperature produced by the three computer models is presented
in Figures \ref{fig:Fine}, \ref{fig:Inter}, and \ref{fig:Corse}.

There is an obvious discontinuity at $x_{1}=0.5$. The accurate model
$\mathscr{C}^{(3)}$ has high frequencies which are not captured by
the lower fidelity models $\mathscr{C}^{(1)}$ and $\mathscr{C}^{(2)}$.
The discrepancy function $\delta_{2}$ varies throughout the input
space, and presents local features such as discrepancies.

For comparison reasons between our proposed method and existing co-kriging
methods, we consider a hierarchically nested design. Hence we compare
the proposed special case BTCK (where augmentation is not needed and
hence dropped out) with the existing GP co-kriging of \citet{Gratiet_SIAM_UQ2013}.
We generate three nested experimental designs for models $\{\mathscr{C}^{(t)}\}_{t=1}^{3}$
according to the condition Latin Hypercube Sampling (cLHS) design
\citep{MINASNY20061378} with sample size $n^{(1)}=150,n^{(2)}=100$
and $n^{(3)}=50$. For prior model, we consider $\phi_{t}|\mathcal{T}\sim0.5\text{G}(1,20)+0.5\text{G}(10,10)$.
The model was trained by running the suggested MCMC sampler for $25000$
iterations and obtaining a sample after thinning the chain by $3$
iterations, and discarding the first $5000$ values as burn in. At
the same datasets, we used the same model parametrization \citet{Gratiet_SIAM_UQ2013}.
For the comparison to be fair, we used the same prior specification
the two approaches.  

The comparison is performed based on the predictive ability of the
procedures. We predict the high-level computer model in a $100\times100$
girded locations and evaluate the mean square prediction error (MSPE)
for both methods.

\begin{figure}[htb!]
\centering \subfloat[\textsf{Bayesian treed co-kriging\label{fig:TGP2pred=00003D00003D00003D00003D00003D00003D00007D}}]{\includegraphics[width=0.45\textwidth]{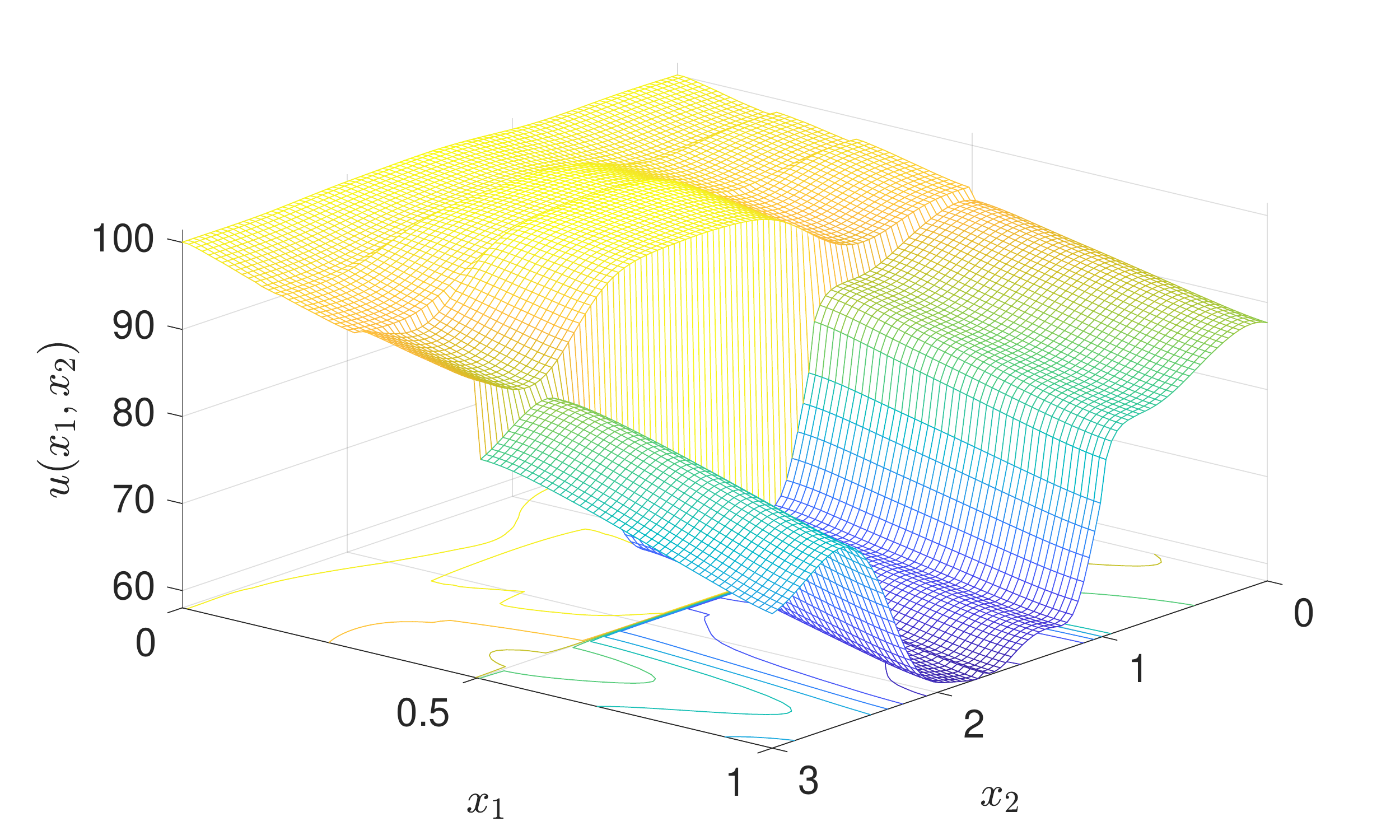}

}\subfloat[\textsf{GP co-kriging \label{fig:GP2pred}}]{\includegraphics[width=0.45\textwidth]{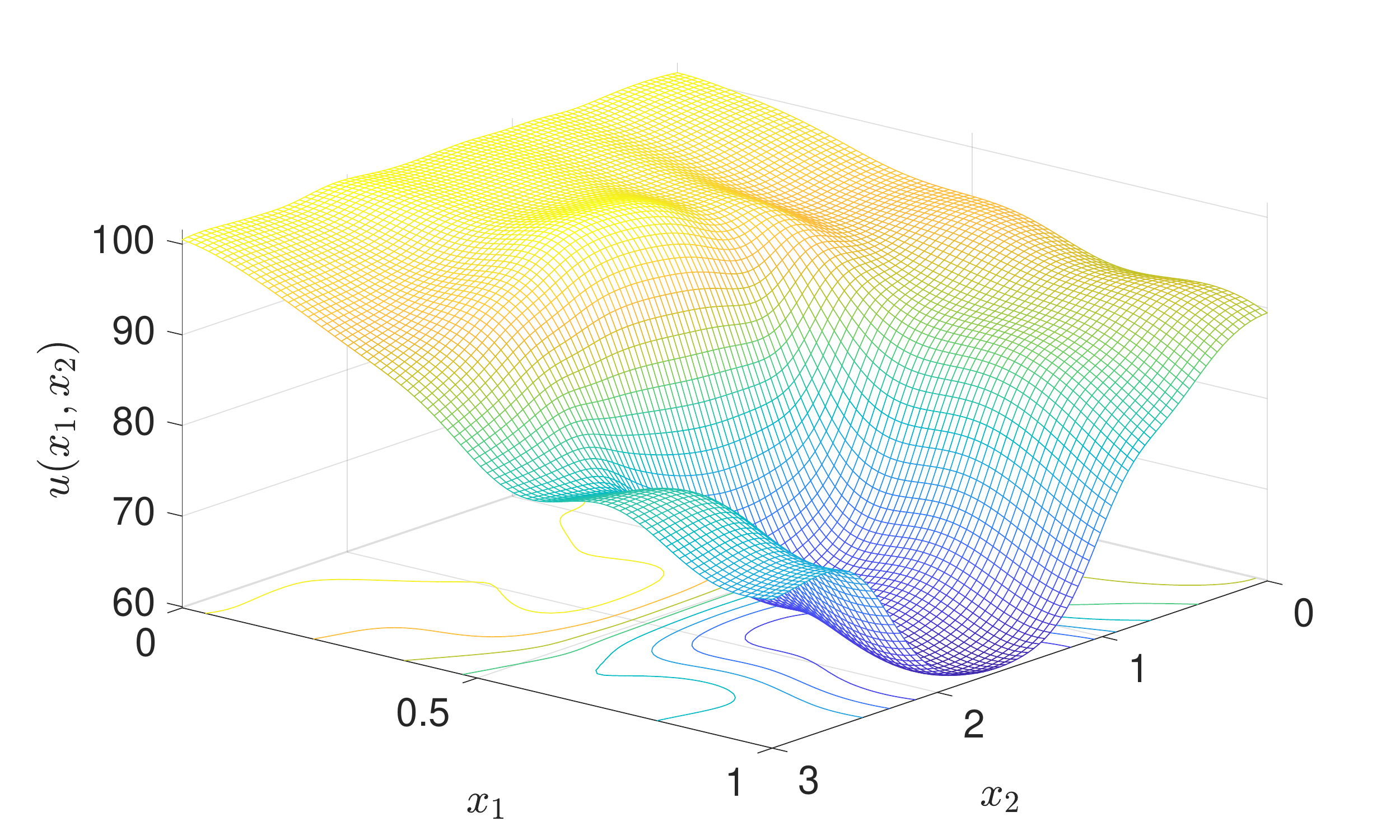}

}

\caption{Prediction mean of the temperature steady state solution for the fine
computer model using two different methods (a) co-kriging GP and (b)
proposed Bayesian treed co-kriging.\label{fig:prediction_plots}}
\end{figure}

In Figures \ref{fig:TGP2pred=00003D00003D00003D00003D00003D00003D00007D}
and \ref{fig:GP2pred}, we present the prediction of the high fidelity
model output for the proposed BTCK and the competitor. 

\begin{figure}[htb!]
\centering \subfloat[\textsf{Scalar Factor $\xi_{1}(x)$ \label{fig:L1L2}}]{\includegraphics[width=0.45\textwidth]{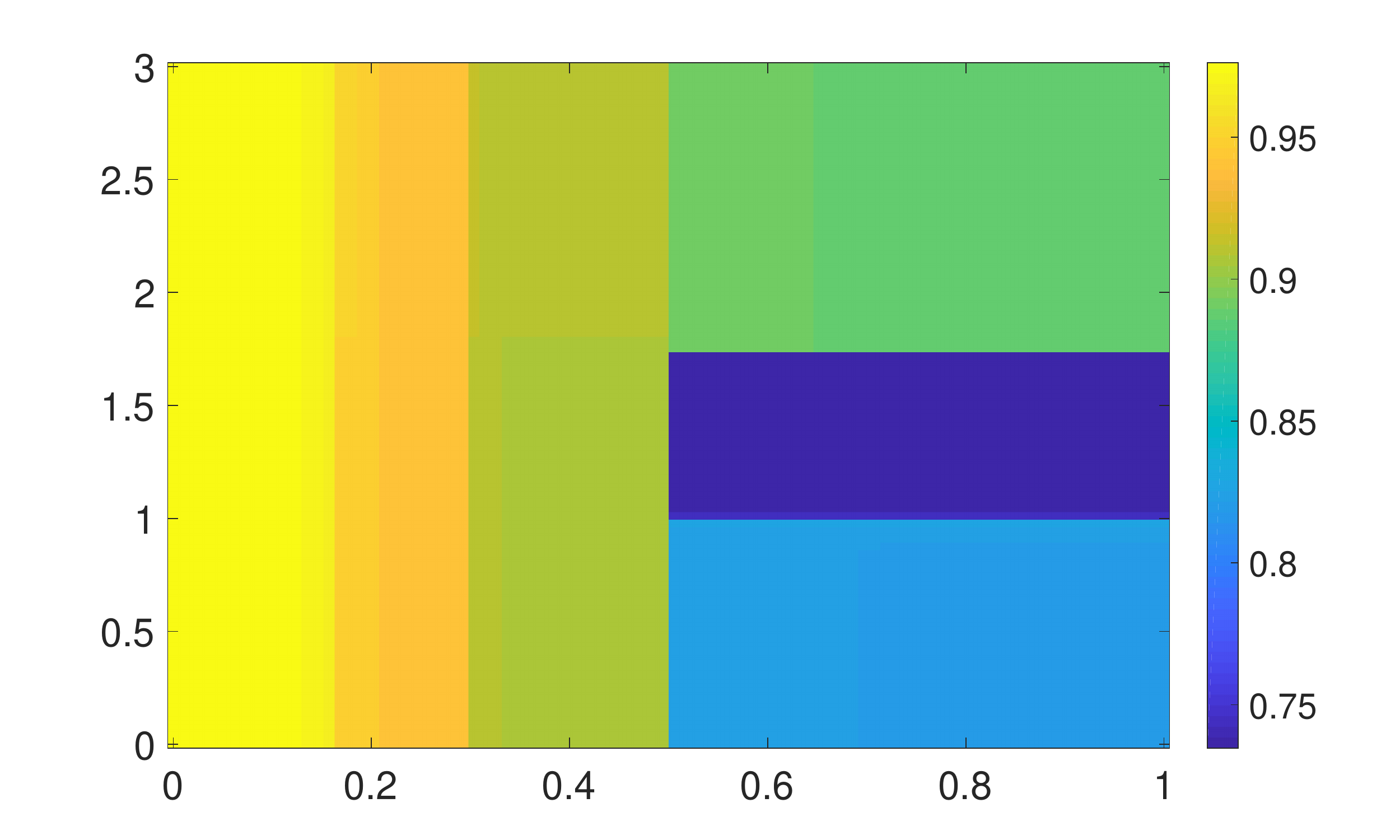}

}\subfloat[\textsf{Scalar Factor$\xi_{2}(x)$ \label{fig:L2L3}}]{\includegraphics[width=0.45\textwidth]{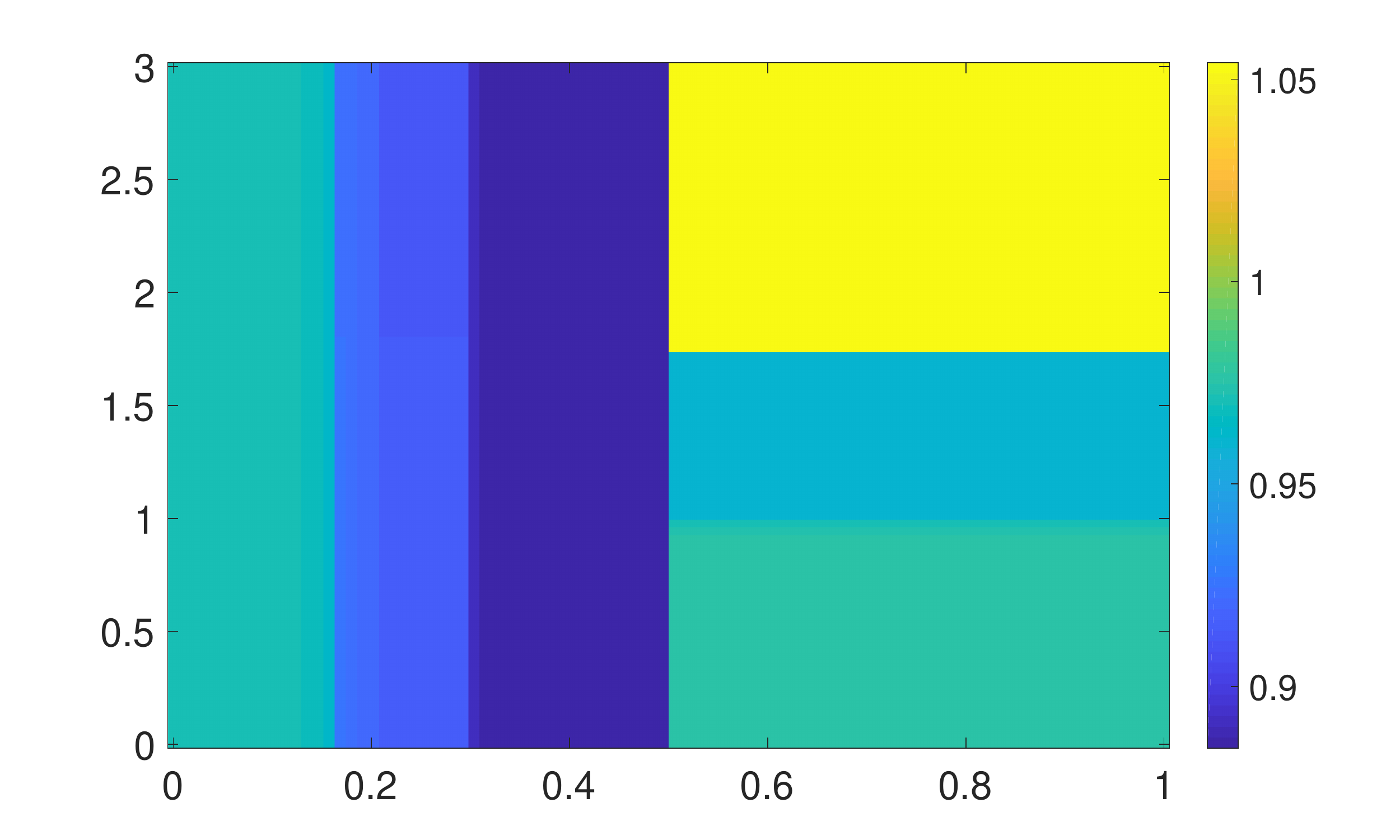}

}

\caption{Estimated mean of the scalar factor between (a) low-level and medium-level
computer models and (b) medium-level and high-level computer models
using the proposed Bayesian treed co-kriging.\label{fig:prediction_plotsEx2-1}}
\end{figure}

We observe that BTCK managed to adequately capture the discontinuity
and the smaller scale variations in the output while the competitor
failed. We speculate that the behavior of the surface produced by
the competitor in Figure \ref{fig:GP2pred} is because the basis expansion
is unable to represent efficiently sudden changes. Moreover, the proposed
BTCK produced a significantly smaller MSPE equal to $1.4613$ compared
to the competitor whose MSPE was $14.1599$. Hence the proposed BTCK
has produced more accurate predictions than the competitor. Also,
ABTCK managed to recover adequately the output function, even though
the design was the same. Figures \ref{fig:L1L2} and \ref{fig:L2L3} demonstrate the estimation
of the scale discrepancy function between models $\mathscr{C}^{1}$
vs. $\mathscr{C}^{2}$ and $\mathscr{C}^{2}$ vs. $\mathscr{C}^{3}$
respectively, as produced by the proposed ABTCK. 

\end{document}